\documentclass[12pt,preprint]{aastex}

\newcommand   {\about} {\mbox{$\sim$}}

\newcommand   {\hh}    {\mbox{H$_2$}}

\newcommand   {\thco}  {\mbox{$^{13}$CO}}

\renewcommand {\deg}   {\mbox{$^\circ$}}
\newcommand   {\arcm}  {\mbox{$^\prime$}}
\newcommand   {\arcs}  {\mbox{$^{\prime\prime}$}}

\newcommand   {\pscm}  {\mbox{cm$^{-2}$}}
\newcommand   {\kms}   {\mbox{km\,s$^{-1}$}}
\renewcommand {\ga}    {\mbox{\rlap{\hbox{\lower5pt\hbox{$\sim$}}}\hbox{$>$}}}
\renewcommand {\la}    {\mbox{\rlap{\hbox{\lower5pt\hbox{$\sim$}}}\hbox{$<$}}}

\newcommand  {\solar}  {\mbox{$_{\odot}$}}

\begin{document}



\def\kms {\hbox{km{\hskip0.1em}s$^{-1}$}} 
\def\ee #1 {\times 10^{#1}}          
\def\ut #1 #2 { \, \mathrm{#1}^{#2}} 
\def\u #1 { \, \mathrm{#1}}          
\def\msol{\hbox{$\hbox{M}_\odot$}}
\def\lsol{\hbox{$\hbox{L}_\odot$}}
\def\kms{km s$^{-1}$}
\def\Blos{B$_{\rm los}$}
\def\etal   {{\it et al. }}                     
\def\psec           {$.\negthinspace^{s}$}
\def\pasec          {$.\negthinspace^{\prime\prime}$}
\def\pdeg           {$.\kern-.25em ^{^\circ}$}
\def\degree{\ifmmode{^\circ} \else{$^\circ$}\fi}
\def\ee #1 {\times 10^{#1}}          
\def\ut #1 #2 { \, \textrm{#1}^{#2}} 
\def\u #1 { \, \textrm{#1}}          

\def\ddeg   {\hbox{$.\!\!^\circ$}}              
\def\deg    {$^{\circ}$}                        
\def\le     {$\leq$}                            
\def\sec    {$^{\rm s}$}                        
\def\msol   {\hbox{$M_\odot$}}                  
\def\i      {\hbox{\it I}}                      
\def\v      {\hbox{\it V}}                      
\def\dasec  {\hbox{$.\!\!^{\prime\prime}$}}     
\def\asec   {$^{\prime\prime}$}                 
\def\dasec  {\hbox{$.\!\!^{\prime\prime}$}}     
\def\dsec   {\hbox{$.\!\!^{\rm s}$}}            
\def\min    {$^{\rm m}$}                        
\def\hour   {$^{\rm h}$}                        
\def\amin   {$^{\prime}$}                       
\def\lsol{\, \hbox{$\hbox{L}_\odot$}}
\def\sec    {$^{\rm s}$}                        
\def\etal   {{\it et al. }}                     

\def\xbar   {\hbox{$\overline{\rm x}$}}         

\slugcomment{Submitted to ApJ}
\shorttitle{}
\shortauthors{}

\title{The Origin of  Diffuse X-ray and $\gamma$-ray Emission from
the Galactic Center Region: Cosmic Ray Particles }
\author{F. Yusef-Zadeh\altaffilmark{1},
M. Muno\altaffilmark{2},
M. Wardle\altaffilmark{3},
D.C. Lis\altaffilmark{4}}

\altaffiltext{1}{Department of Physics and Astronomy,
Northwestern University, Evanston, Il. 60208
(zadeh@northwestern.edu)}
\altaffiltext{2}{Dept Physics and Astronomy, University of California, Los 
Angeles, Box 951547, Los Angeles, CA 90095   
(mmuno@astro.ucla.edu)}
\altaffiltext{3}{Department of Physics, Macquarie University, Sydney NSW 2109,
Australia (wardle@physics.mq.edu.au)}
\altaffiltext{4}{California Institute of Technology,
MC 320-47, Pasadena, CA 91125 
(dcl@caltech.edu)}

\begin{abstract} 
The inner couple hundred pcs of our Galaxy is characterized by significant amount of synchrotron-emitting 
gas, which appears to co-exist with a large reservoir of molecular gas.  Many of the best studied sources in 
this region exhibit a mixture of 6.4 keV Fe K$\alpha$ emission, molecular line emission and nonthermal radio 
continuum radiation.  The spatial correlation between  fluorescent~Fe K-$\alpha$ line  emission at 6.4 keV and 
molecular line emission~  from Galactic center molecular clouds has been explained as reflected X-rays from a 
past outburst of Sgr~A*. Here we present multi-wavelength study of a representative Galactic center cloud Sgr 
C using {\it Chandra}, {\it VLA} and {\it FCRAO}.  We note a correlation between the nonthermal radio 
filaments in Sgr C and the X-ray features, suggesting that the two are related.  This correlation, when 
combined with the distribution of molecular gas suggests against the
irradiation of Sgr~C by Sgr A*.  Instead, we 
account for this distribution in terms of the impact of the relativistic particles from local (nonthermal 
filaments) and extended  sources with diffuse neutral gas producing both a nonthermal bremsstrahlung X-ray 
continuum emission, as well as diffuse 6.4 keV line emission.  The production rate of Fe K$\alpha$ photons 
associated with the injection of electrons into a cloud as a function of column density is calculated.  The 
required energy density of low-energy cosmic rays associated with the synchrotron emitting radio filaments or 
extended features is estimated to be in the range between 20 and $\sim10^3$ eV cm$^{-3}$ for Sgr C, Sgr B1, 
Sgr B2, and ``the 45 and -30 \kms'' clouds.  We also generalize this idea to explain the cosmic-ray heating 
of molecular gas, the  interstellar cosmic ray ionization,  the  pervasive 
production 
of 
diffuse K${\alpha}$ line and TeV emission from the Galactic center molecular clouds.  In particular, we 
suggest that Inverse Compton scattering of the sub-millimeter radiation from dust by relativistic electrons 
may contribute substantially to the large-scale diffuse TeV emission observed towards the central regions of 
the Galaxy.
\end{abstract}
\section{Introduction}
\label{introduction} 

It is well recognized that the Galactic center region hosts several sources of 
energetic activity in the form 
of nonthermal linear filaments,  supernova remnants and colliding winds of massive stars. 
This region 
is also 
recognized to be the 
site of massive molecular clouds with pockets of on-going and past massive star formation. 
The most prominent of these clouds are associated with Sgr B2, Sgr C, as well as the clouds 
associated with radio continuum arc and Sgr A.  
Sgr~C or G359.43--0.09 is  one of the best examples  of a massive star-forming region which 
hosts 
prominent thermal and nonthermal radio continuum, millimeter,  
sub-millimeter and X-ray sources 
in the 
Galactic center region. 
This massive star forming region is identified by its H\,{\sc ii} region, a dense molecular cloud and 
nonthermal radio filaments. This source is also known as an X-ray source based on ASCA 
observations (Murakami et al. 2001a) showing the evidence of iron K-shell fluorescence 
(K${\alpha}$) line emission. To probe the nature of high-energy activity in the Galactic 
center region, we have selected to study Sgr~C in detail, as well as 
star forming regions near the continuum arc ($l \sim 0.2^{\circ}$) and the Sgr B complex. 

Sgr~C consists of a thermal radio continuum, a far-IR and a 
sub-millimeter source, as well as a synchrotron emitting source 
associated with magnetized filaments (Odenwald \& Fazio 1984; Caswell \& 
Haynes 1987;  
Liszt 1992; Tsuboi et al. 1991; Lis \& Carlstrom 1994). 
The  Sgr~C H\,{\sc ii} complex  
is thought to be powered by a single 
O4  ZAMS star. The  molecular mass of Sgr C is 
estimated to be 
$\sim5-6\times10^5$ \msol with column 
density of $N_{\rm H_2}\sim10^{23}$ cm$^{-2}$, based on 
sub-millimeter and $^{13}$CO observations (Liszt \& Spiker 
1995; Lis \& Carlstrom 1994).
Recent radio continuum survey of the Galactic center region has 
identified additional nonthermal filaments distributed in 
the immediate vicinity of the Sgr C H\,{\sc ii} region (Nord et al. 2004; 
Yusef-Zadeh, Hewitt \& Cotton 2004). 
The main nonthermal radio 
filament (NRF)  appears to end abruptly inside the Sgr~C 
molecular H\,{\sc ii} complex M359.5--0.15 with a velocity of --65 \kms 
(Liszt \& Spiker 1995).  Similar radio recombination line velocity has also been 
detected from the Sgr C H\,{\sc ii} region
(Anantharamaiah \& Yusef-Zadeh 1989). 
 HI absorption measurements toward 
the nonthermal filament and the H\,{\sc ii} region constrain the 
distance to both sources within the inner 200 pc of the Galactic 
center. However, Roy (2003)  suggests that the Sgr~C filament and H\,{\sc ii} 
region   are 
two separate objects and may not be interacting with each other (Roy 
2003).

The ASCA measurements show the presence of strong fluorescent 
6.4 keV line emission from neutral iron gas implying H$_2$ 
column density similar to that of molecular line observations 
(Murakami et al. 2001a). The nature of the 6.4 keV line emission 
 has been considered to 
be due to the irradiation of the Sgr~C molecular cloud 
originated by past energetic outburst from Sgr~A* (Murakami et 
al. 2001a). Similar mechanism  has been postulated to explain the 
origin of the strong 6.4 keV line emission detected from other 
Galactic center molecular clouds (e.g., Sgr~B2) by either Sgr A* or 
from the interaction of the Sgr A East and ``the 50 \kms cloud''
(Koyama et al. 1996;  Murakami et al. 2000, 2001b; 
Revnivtsev et al. 2004; Fryer et al. 2006). 
An alternative explanation to the X-ray 
reflection nebula model has  accounted for the origin of X-ray emission  
in terms of  the impact of low-energy cosmic 
ray particles with neutral gas associated with the 
M0.11--0.08 molecular cloud (Yusef-Zadeh, Law \& Wardle 
2002).  
These authors speculated that the same process 
could also explain the origin of the 6.4 keV line emission from 
 other  Galactic center 
molecular clouds.  

The structure of this paper is as follows. In $\S$2, we re-analyze the archival {\it 
Chandra} observations of the Galactic center region followed by the analysis of new 
millimeter line and radio continuum data using FCRAO and VLA.  In $\S3$, we report 
the discovery of X-ray filamentary structures in Sgr C. The morphology of X-ray and 
nonthermal radio continuum features suggest that many of the K$\alpha$ 6.4 keV line 
features lie near the nonthermal radio filaments at 20 and 90~cm wavelengths. We also 
examine the morphology of \thco\, and Fe K$\alpha$ lines and estimate column 
densities using millimeter and X-ray techniques. 
 We  present spectral index determination in $\S3.1.5$   and 
polarization measurements of the brightest filament in Sgr C at radio wavelengths in the Appendix. 
The synchrotron parameters of the Sgr C filament is critical 
in support of our picture that the low-energy component of the synchrotron emitting 
particles penetrate molecular clouds. 
We then compare the distribution of 6.4 keV 
and TeV emission from prominent galactic center molecular clouds. The overall 
correlations of K$\alpha$ line emission, molecular line emission from the ``45 and 
-30 \kms\ clouds'', TeV emission and nonthermal radio continuum features are 
described. In $\S$4, we argue in support of enhanced cosmic rays in the central 
region of the Galaxy and compare the irradiation model versus the low-energy cosmic 
ray model to account for the origin of the 6.4 keV emission. We then describe a 
quantitative mode of injection of cosmic rays into molecular clouds and the 
subsequent heating of gas before this model is applied to the Galactic center 
molecular clouds. Unlike the irradiation model, the picture that is presented here 
can naturally explain 
the high temperature of molecular gas in the galactic center region. 
Lastly, we 
investigate the origin of
 diffuse TeV emission from Galactic center clouds in terms of the 
inverse Compton scattering of dust emission by the high energy component of the 
cosmic rays distributed throughout this unique region of the Galaxy.


\section{Data Reductions}
\subsection{X-ray Analysis}

Sgr~C was observed with the {\it Chandra} Advanced CCD Imaging 
Spectrometer (ACIS-I) (Weisskopf et al. 2002)
as part of the shallow 
Galactic center survey   (Wang et al. 2002)
with two overlapping 11 
ks exposures on 2001 July 20 (sequences 2270 and 2272). We 
reduced the observation using standard tools that are part of 
CIAO version 3.2. We first created a cleaned event list from the 
ACIS-I array for each observation. We corrected the pulse 
heights of the events for position-dependent charge-transfer 
inefficiency  (Townsley et al. 2002)
and excluded events that did 
not pass the standard ASCA grade filters and {\it Chandra} X-ray 
center (CXC) good-time filters.  We searched for intervals 
during which the background rate flared to $\ge 3\sigma$ above 
the mean level, and found none. Our goal was to study the 
properties of diffuse features, so we removed events within 
circles enclosing at least 92\% of the point-spread function 
that were centered at the locations of the point-like X-ray 
sources,  as identified  by Muno et al. (2006). Finally, we re-projected 
the events from 
each observation onto a fixed tangent point (the location of Sgr~A*), so 
that 
they could be combined to form composite images.

We then generated images of the continuum emission between 4--8 
keV, excluding the 6.10--7.15 keV energy range containing iron 
lines, with a resolution of 9\arcsec. 
In addition to the continuum images, we also
made  the Fe K-$\alpha$ flux  and equivalent width  images. 
Both the
continuum and line flux images were made the same way. 
For display purposes, the 
holes left in the image when we excluded point sources were 
filled with pixel values drawn from a Poisson distribution that 
had the mean value of the pixels in a surrounding annulus. We 
repeated the same process on an image of an exposure image 
generated with standard CIAO tools, and then divided the count 
image by the resulting exposure image in order to create a flux 
image  of the diffuse emission. Finally, we present two different types of images, 
one of which is
adaptively smoothed 
using the CIAO tool csmooth, whereas the other is  convolved
 with 30\arcs\
Gaussians.


Images  of the
equivalent widths of the low-ionization 
6.4 keV line of Fe were constructed
using the 
techniques  described by Park et al. (2004).
 Adaptively-smoothed images  of the 
diffuse line emission were generated in the same manner as the 
continuum image, using the 6.25--6.50 keV band for Fe K-$\alpha$.
 The continuum under each 
line was computed based on adaptively-smoothed images of the 
flux in the 5.0--6.1 keV and 7.15--7.30 keV energy bands. We assumed 
that the flux in each continuum band ($F_{\rm band}$) could be 
described as a power law, so that the normalization ($N$) and 
slope ($\Gamma$) of the power law could be computed from 
\begin{equation} F_{\rm band} = {{NE_{\rm low}^{-\Gamma+1} - 
NE_{\rm high}^{-\Gamma+1}}\over{\Gamma - 1}}. \end{equation} 
Using the fluxes in both continuum bands, the above equation was 
solved for $N$ and $\Gamma$ using Newton's algorithm, and the 
parameters were used to estimate the continuum contribution to 
the line emission images. To derive the equivalent width images, we 
subtracted the estimated total continuum flux from the line image, 
and then divided the line image by the continuum flux density at 
the centroid of the line (6.4 keV).
We caution that we have neglected the cosmic-ray background
in generating these maps, which could account as much as  $\sim$40 \% of the 
events in the 6--7 keV band, and consequently biases any estimate of
the
equivalent width. The assumption of a power law spectrum also
introduces
a small systematic bias in these maps. We have not attempted to
correct
these effects, because they are only used to search for regions of 
enhanced iron emission.

In order to confirm the properties suggested by the images  of 
the 
diffuse line and continuum emission, we examined the spectra of 
small regions of the image. For each observation, counts were 
extracted from each region of interest, and then binned as a 
function of pulse-height to create spectra. We obtained the 
response functions from Townsley et al. (2002)
and averaged them 
weighted by the number of counts received by the relevant 
portions of the detectors. Effective area functions for the 
spectra of each observation were generated using {\tt mkwarf}, 
and averaged weighted by the counts in each observation. The 
background was estimated from the spectrum of the particles that 
impacted the detectors during a 50~ks observation taken with the 
ACIS-I stowed out of the focal plane of the mirror assembly. 


\subsection{Millimeter Line Analysis}

The \thco~(1--0) data presented here were taken in 1993 April using
the 15-element QUARRY focal plane array mounted at the Cassegrain
focus of the 14-meter FCRAO telescope. Each pixel of the receiver
employed a cooled Schottky diode mixer and 1.3--1.7 GHz HEMT IF
amplifier. There was a common quasioptical single sideband filter with
cooled image termination Erickson et al. (1992). The backend 
was a
wideband filterbank spectrometer having 32 channels with 5~MHz
(13.6~\kms) resolution for each pixel of the array. Typical system
temperatures referred to above the earth's atmosphere were \about
800--1200~K, the FWHM beam size was \about 50\arcs, and the main beam
efficiency \about 40\%. The \thco\ spectra were taken at 720
positions, with approximately one beam spacing, covering an area of
$25\arcm \times 20$\arcm. The raw data were converted to CLASS format and 
the subsequent data
reduction and analysis were carried out using the IRAM GILDAS software
package.

\subsection{Radio Continuum  Analysis}
The 20~cm images presented here are based on observations with 
the Very Large Array (VLA)
of the National Radio Astronomy Observatory\footnote{The National
Radio Astronomy Observatory is a facility of the National Science
Foundation, operated under a cooperative agreement by Associated
Universities, Inc.} observations that were carried out using B, C and
D configurations.  Details of data reductions are described in
Yusef-Zadeh, Hewitt, \& Cotton (2004).  To measure the spectral index
distribution, we compared the 20~cm image with a 90~cm image which was
also based on multi-configuration observations with the VLA (Nord et
al.  2004).  The spatial frequency coverage of both 20 and 90~cm data
were similar to each other so an accurate spectral index distribution
could be determined.  We convolved both images to a Gaussian having a
FWHM=$12.6''\times12.6''$ before the spectral index image was
constructed in AIPS. The rms noise for the two images are 3.6 and 2.5
mJy beam$^{-1}$ at 90 and 20~cm, respectively.

To search for polarized emission from Sgr C, we also reduced radio continuum 
observations of Sgr C at 3.6 and 2~cm 
which were observed in the DnC hybrid array configuration of the VLA in 
June 19, 1988. Standard calibration was done using NRAO530 and 3C286  
as the phase and flux calibrators,  respectively.  The phase center
for both pointings at 3.6 and 2~cm wavelengths is centered  at 
$l$=359$^{\circ} 27' 14.04''$, $b$=--2$' 10.3''$.  Based on 3.6 and 2~cm, we also 
placed a limit on  the spectral index of the nonthermal filaments in Sgr C. 

\section{Results} 

We have compared the spatial distribution of X-ray, nonthermal radio continuum, molecular line
and  TeV emission and find a 
rough correspondence that suggests the the they  are related with each other. In particular, the 
analysis of the diffuse X-ray, molecular line and nonthermal radio continuum emission 
derive the iron flux, the column depth of molecular gas and the nonthermal particle flux. 
These  basic observables are then used to measure how cosmic rays interact with neutral material.

\subsection{The Sgr~C Molecular Complex} 
\subsubsection{Diffuse X-ray 
Emission}

Figure 1a shows contours 
of X-ray 
continuum emission from Sgr~C
between 2-6 keV. We note the evidence for several extended X-ray
features in this H\,{\sc ii} complex.  Diffuse features run vertically
perpendicular to the Galactic plane, one of which is G359.42--0.12 with
an extent of about $>4'$ running to the south and the other is
G359.45--0.07 with similar extent but appears clumpy as it runs to the
north.  Three relatively isolated and compact features are also found
near G359.4--0.07, G359.32--0.16 and G359.46--0.15 with an extent of
$\sim$1\arcmin.  These individual X-ray features will be discussed in
$\S$3.3 and $\S$3.4.

We extracted an average X-ray spectrum 
of Sgr~C from a region that enclosed the brightest parts of the diffuse 
and compact  
emission, as noted in Figure 1a,  in the smoothed line images. 
 We accounted for the background using the 
particle spectrum before we made the  X-ray spectrum, as   shown in 
Figure 1b. The extracted region is 
shown as an ellipse in a grayscale  6.4 keV EW image in Figure 2.  
The resulting spectrum of Sgr~C, looks very similar to 
that of the diffuse emission analyzed by Muno et al. (2004a).  
The emission spectrum was modeled with  
 a two-temperature  
plasma with and without a power-law  component using  the {\tt XSPEC} 
spectral 
modeling package (Arnaud 1996). 
We assumed that each component was in
collisional
ionization equilibrium (the {\tt apec} model in {\tt XSPEC}),
although the two components were not physically
coupled (e.g., by conduction). We assumed that both components were
absorbed by the same interstellar gas and dust, which we took to be
distributed non-uniformly
along the line of sight and across the image such that its effect
could be described mathematically as
\begin{equation}
e^{-\sigma(E)N_{\rm H}}([1-f] + fe^{-\sigma(E)N_{\rm pc,H}}).
\label{eq:abs}
\end{equation}
Here $\sigma(E)$ is the energy-dependent absorption cross-section,
$N_{\rm H}$ is the absorption column affecting the entire plasma,
$N_{\rm pc,H}$ is the column affecting part of the column, and $f$ is
the fraction of the plasma affected by $N_{\rm pc,H}$.
The parameters of the best-fit model are listed in
Table~\ref{tab:spec}.

The spectrum of the X-ray emission from Sgr C resembles that toward the
inner 20 pc of the Galaxy, in that it can be modeled as plasma
components
with $kT_{\rm s}$$\approx$0.5 and $kT_{\rm h}$$\approx$8 keV
\citep[][]{mun04a}.\, It is highly absorbed, with the column density
toward $>$97\% of the region reaching $10^{23}$ cm$^{-2}$, which is
comparable to the total column depth of the molecular cloud.
The derived emission measure of the soft plasma, $\int n_e n_{\rm H}
dl$
(Table~\ref{tab:spec}), is highly dependent on the absorption column,
because
a $\approx$0.5 keV plasma emits $<$1\% of its flux in the 2--8 keV band
and
photons below 2 keV are absorbed by the interstellar medium. Therefore,
there
is a factor of $\sim$100 systematic uncertainty in the emission measure
of the
soft plasma, and our results cannot be compared easily to those using
different
absorption models. 
The emission measure of the hard plasma is less affected by absorption.
We find a value of 1.0$\pm$0.1 cm$^{-6}$ pc toward Sgr C, which is only
a
factor of two less than the average value of the inner 20 pc of the
Galaxy
\citep{mun04a}.\footnote{We note that in Table 3 of
\citet{mun04a}, the emission measure is erroneously listed as being
in units of $10^{-4}$ cm$^{-6}$ pc. It should actually be cm$^{-6}$ pc,
as
in Table~\ref{tab:spec} of this paper.} Likewise,
the intensity of the Fe K-alpha emission toward Sgr C is similar to
that
in the central 20 pc of the Galaxy,
$\sim 1.2\pm0.1 \times 10^{-6}$ ph cm$^{-2}$ s$^{-1}$ arcmin$^{-2}$.
The equivalent width of the Fe K-$\alpha$ line is 460$\pm$100 eV.

In the case where the spectrum was modeled by the inclusion of 
a power-law 
component, the fit is no better with a power law 
than without one. If we assume that the hot plasma producing the 6.7 keV 
iron 
lines  has solar abundance, then up to 20\% of the total de-absorbed 2-8 
keV flux could be produced by a power law with the photon 
index $\Gamma$   between 1 and 2.  If we assume that the plasma has 
twice-solar iron  abundance, then up to 45\%~of the flux could be 
produced by a non-thermal plasma. If larger abundances are assumed, as 
some Galactic center diffuse features such as the Arches and Quintuplet 
clusters indicate (e.g., Yusef-Zadeh et al. 2002; Law \& Yusef-Zadeh 
2004; Wang, Dong, \& Lang 2006), it would imply that the line  emission is 
accompanied by very 
little thermal continuum, so that the  power law can contribute yet more 
to the spectrum. 
\clearpage
\begin{deluxetable}{lc}
\tablecolumns{2}
\tablewidth{0pc}
\tablecaption{Spectrum of the Diffuse X-ray Emission from Sgr C\label{tab:spec}}
\tablehead{
\colhead{Parameter} & \colhead{Value} \\
}
\startdata
$N_{{\rm H},1}$ ($10^{22}$ cm$^{-2}$) & $ 4.0_{- 0.1}^{+ 0.3}$ \\
$N_{{\rm H},{\rm pc},1}$ ($10^{22}$ cm$^{-2}$) & $10.7_{- 0.7}^{+ 0.5}$\\
$f_{{\rm pc},1}$ & $>$0.97 \\
$kT_1$ (keV) & 0.5$\pm$0.1 \\
$Z_1$ & 0.3$\pm$0.1 \\
$K_{\rm em,1}$ (cm$^{-6}$ pc) & $231.5_{-70.7}^{+79.2}$ \\
$kT_2$ (keV) & $7.6_{-0.8}^{+1.0}$ \\
$Z_2$ & 0.6$\pm$0.1 \\
$K_{\rm em,2}$ (cm$^{-6}$ pc) & 1.0$\pm$0.1 \\
$F_{\rm Fe K}\alpha$ ($10^{-7}$ ph cm$^{-2}$ s$^{-1}$ arcmin$^{-1}$) &12$\pm$1 \\
$\chi^2/\nu$ & 335/325 \\ 
$F_{\rm X,s}$ ($10^{-14}$ erg cm$^{-2}$ s$^{-1}$ arcmin$^{-2}$) & 3 \\
$F_{\rm X,h}$ ($10^{-14}$ erg cm$^{-2}$ s$^{-1}$ arcmin$^{-2}$) & 9 \\
$L_{\rm X,s}$ ($10^{33}$  erg s$^{-1}$ arcmin$^{-2}$) & 3 \\
$L_{\rm X,h}$ ($10^{33}$ erg s$^{-1}$ arcmin$^{-2}$) & 2
\enddata
\tablenotetext{a}{Parameter was not allowed to vary.}
\tablecomments{Uncertainties are 1$\sigma$, computed using
$\Delta \chi^2 = 1.0$. The emission measure is defined as
$K_{\rm em} = \int n_e n_{\rm H} dl$, where $dl$ is the integral over
the length in parsecs. Fluxes are observed values in the 2--8 keV band,
and luminosities are computed from the de-reddened flux in the 2--8 keV
band. The soft component emits $<$1\% of its flux in the
2--8 keV band, so its emission measure could be lower by a factor
of $\approx$100 if it is produced in a region with a lower column
density
than the hard emission.}
\end{deluxetable}
\clearpage
\subsubsection{X-ray and 20-cm Continuum Emission}

Figure 3a,b show contours of X-ray continuum emission between 2-6 keV 
superimposed on two  grayscale renditions of continuum emission 
 from   Sgr~C at 20~cm. The 20~cm image shows 
the 
circular Sgr~C H\,{\sc ii} region, as well as the prominent vertical structure 
which consists of bright  nonthermal filaments running vertically toward 
more 
positive latitudes in the direction away from  the Galactic plane.
Figure 3c  shows the same contours superimposed on a 90~cm image of the Sgr C  region 
to present  the relative distribution of nonthermal continuum and X-ray emission. 
Due to their steep spectrum, 90~cm continuum emission is generally  a good tracer of nonthermal 
emission from synchrotron  sources (Nord et al. 2004). We describe individual nonthermal radio 
filaments and argue that each radio filament has X-ray correspondence. The peaks of radio
and X-ray features appear to be  displaced between 15$''$ and 60$''$ with 
respect to each other. 
Five   X-ray features are labeled in Figure 1a, all of which 
appear to be  spatially correlated with the nonthermal radio 
continuum  features, as each described below. 

\begin{description}
\item[G359.45--0.07]
One of the diffuse but clumpy X-ray features, G359.45--0.07, as labeled
in Figure 1a, which runs along the brightest portion of the nonthermal
radio filament of Sgr C. This nonthermal radio counterpart, labeled as
RF-C1 (G359.45--0.06) in Yusef-Zadeh et al.  (2004), is prominently
detected at 90~cm (Nord et al.  2004).

\item[G359.42--0.12]
Another X-ray feature which appears 
to have a radio correspondence  is 
G359.42--0.12  running  parallel to   a new vertical radio continuum  
feature. This 
 new  faint radio feature, as noted best in  Figure 3b,  
extents  for   8$'$ with a typical 
surface brightness of 
$\sim$10 mJy beam$^{-1}$. 
This   radio continuum feature 
at $l$=359$^{\circ} 26'$  
runs  parallel to the 
eastern edge of 
the elongated X-ray feature G359.42--0.12. The centroid of the radio 
continuum and the X-ray continuum features is displaced by $\sim1'$ in 
east-west direction.  

\item[G359.40--0.07]
The third   X-ray feature that appears to have 
a radio correspondence  is G359.40--0.007 (Figs 3a,c). This X-ray feature 
lies at the northeastern edge of a  filamentary 
radio feature which has been identified 
earlier at 20~cm (see RF-C14 (G359.32--0.06) in Yusef-Zadeh et al. 2004) and 
at 90~cm  (Nord et al. 2004). 

\item[G359.46--0.15]
Another  X-ray feature G359.46--0.15 is  noted at the southwestern  
edge of  a long filamentary structure identified  as RF-C4 (G359.49--0.12) 
at 20~cm (Yusef-Zadeh et al. 2004). Figure 3b shows the relative location 
of this faint  X-ray emission with respect to the filament. An H\,{\sc ii} region 
(G359.47--0.17) lies to the south of the X-ray feature. 

\item[G359.32--0.16]
Lastly, we  notice in Figures 3a-c, a compact X-ray feature 
G359.32--0.16 which 
coincides with the brightest component of a nonthermal radio filament. 
This filament  
runs  diagonally with a position angle of $\sim60^{\circ}$ and  is identified 
at 20~cm as RF-C13 in Yusef-Zadeh et al. (2004) and at 90~cm  (Nord et al. 
2004).   High resolution radio continuum images show that the radio 
filament
consists of multiple linear components 
running parallel to each other but peaking  at the location 
of the X-ray feature G359.32--0.16. 
Figure 3d shows a close-up view of G359.32--0.16 where contours of 
X-ray emission are superimposed on a grayscale 20~cm continuum image. 
The distribution of X-ray emission is elongated along 
the direction of two parallel 
nonthermal radio filaments,  suggesting that the X-ray and radio
features could  be associated with each other. Like other X-ray features 
that have radio continuum correspondence, a displacement of 
$\sim15''$ is 
noted 
between the peaks of X-ray and radio emission. 
The displacement between nonthermal radio filaments 
and X-ray features is also seen in high-resolution images
 G0.11--0.08 (Yusef-Zadeh, Law, \& Wardle 2002). 

We 
extracted events from the isolated continuum X-ray feature G359.32--0.16.  
The events were selected from a 35.4$'' \times 57.6''$ ellipse centered 
at  $l$=359.43382$^{\circ}$, $b$=--0.09655$^{\circ}$ with radius 4.15\arcmin\ and  
5.96\arcmin\, 
respectively, with a position angle (PA=38.6$^{\circ}$). 
The background was estimated using a local background. 
After accounting for the background, the region contains only 45$\pm$16 
net photons (90\% uncertainty) in the {\it Chandra} bandpass, all of which 
lie between 3.3-8.0 keV; in the 0.5-2.0 keV band, there are $<$7 net counts, 
and in the 2.0-3.3 keV band there are $<$11 net counts. This does not 
provide enough signal to model the spectrum in detail. Therefore, 
following Muno et al. (2004b), we calculated a net photon flux of 
$1\times10^{-7}$ photon cm$^{-2}$ s$^{-1}$, and a hardness ratio 
$(h-s)/(h+s) = 0.2 \pm 0.4$, where $h$ is the number of counts in the 
4.7--8.0 keV band and $s$ is the 3.3--4.7 keV band. If the absorption 
column to the source is $6\times10^{22}$ cm$^{-2}$, then we can use 
Figure~13 in Muno et al. (2004b) to estimate that the spectrum is 
consistent with a $\Gamma$=1 power law with a luminosity is 
$1\times10^{33}$ erg s$^{-1}$. Unlike other radio filaments in Sgr C, 
G359.32--0.16 does not seem to have 
a 6.4 keV correspondence, thus it is possible that this X-ray feature  is 
an X-ray  synchrotron 
source similar to other X-ray/radio nonthermal filaments that have been 
detected  in the 
Galactic center region (Sakano et al. 2003; Lu, Wang \& Lang 2003; 
Yusef-Zadeh et  al. 2003).  
\end{description}

In addition to radio filaments described above, Sgr C is surrounded by 
a number of additional weaker  
radio filaments that have been identified at 20 and 90~cm wavelengths (Nord et al. 2004; 
Yusef-Zadeh, Hewitt \& Cotton 2004). 
We predict that  a  large-scale  sensitive X-ray observations of this
region with {\it Chandra} and 
{\it XMM-Newton}
X-ray Observatories  should 
detect additional X-ray correspondence  to radio features. 


\subsubsection{$^{13}$CO Line Emission}

The \thco~(1--0) emission in the vicinity of Sgr~C spans a wide range
of velocities, from approximately --200 to +200~\kms. Velocity channel
maps  in the --172 to 86~\kms\ range, where the brightest emission
occurs, are shown in celestial coordinates in Figure 4. A number of 
distinct
velocity components can be identified, some with velocities forbidden
by the galactic rotation. The emission associated with the Sgr~C
molecular cloud is seen in the central part of the map in the --77 to
--36~\kms\ channels. In addition, an elongated  feature approximately
parallel to the galactic plane is seen to the west of Sgr~C at lower
velocities (below --90~\kms). Another elongated feature, roughly
perpendicular to the galactic plane, is seen in the north-east part
of the map at velocities above --25~\kms.

Molecular column densities can be computed using eq. (8) of
Lis \& Goldsmith (1991) for a linear molecule:

\begin{equation}
N_{tot} = {{4.0 \times 10^{12}\, {\rm cm^{-2}}} \over {J^2\,
  \mu^2[{\rm D}]\, B[{\rm K}]}} \, Z \, \exp \left( {E_J[{\rm K}] \over
  T_{ex}} \right) \times { 1 \over \eta } \int T_A^* dv \, [{\rm K\, kms^{-1}}]
\end{equation}

\noindent In our case, $J=1$, the dipole moment $\mu = 0.112$~D, the
 rotational constant $B = 2.64$~K, the partition function $Z = T_{ex}
 / B$, the upper level energy $E_1 = 2 B$, and the main beam
 efficiency $\eta = 0.4$. Therefore

\begin{equation}
N_{tot} = 1.14 \times 10^{14} \, {\rm cm^{-2}} ~ T_{ex} \, 
\exp \left( { 5.28 \over T_{ex} } \right) \int T_A^* dv \, [{\rm K\, kms^{-1}}]
\end{equation}

\noindent The assumption of optically thin emission is justified given
the relatively low \thco~(1--0) line intensities compared to the peak
line intensity observed in Sgr~B2 (Lis \& Goldsmith 1989). The dependence of the
molecular column density on the excitation temperature is relatively
mild, as the factor $T_{ex} \exp (5.28 / T_{ex})$ varies between 21
and 46 for $T_{ex}$ between 15 and 40~K. In the subsequent analysis we
assume $T_{ex} = 25$~K (the cold component
in the two-component gas model of H\"uttemeister et al. 1993), and the fractional \thco\ abundance 
$X =10^{-6}$, as determined for Sgr~B2 (Lis \& Goldsmith 1991), which gives

\begin{equation}
N_{\rm H_2} =  3.5 \times 10^{21} ~ {\rm cm^{-2}} ~ \int T_A^* dv \, [{\rm
    K\, kms^{-1}}] 
\end{equation}

\noindent The uncertainty in the molecular column densities is of
order 40\%, taking into account the uncerinties in the calibration and
the excitation temperature.

The peak \thco~(1--0) intensity integrated over the --200
to 200~\kms\ velocity range is 160~K\,\kms\ and the average value over
the area covered by our observations is 84~K\,~\kms. The corresponding
\hh\ column densities are $5.6 \times 10^{23}$ and $2.9 \times
10^{23}$~\pscm, respectively. Limiting the velocity range to --85 to
--30~\kms, corresponding to the Sgr~C cloud, results in the peak and
average integrated line intensities of 80 and 22.3~K\,\kms,
respectively. The corresponding \hh\ column densities are $2.8 \times
10^{23}$ and $8 \times 10^{22}$~\pscm, respectively. 
Assuming a
distance of 8.5~kpc, the total \hh\ mass contained within our map is
$1.5 \times 10^6$~M\solar\ in the --200 to 200~\kms\ velocity range and
$4 \times 10^5$~M\solar\ in the --85 to --30~\kms\ range. These
values are much lower compared to the molecular mass of the Sgr~B2
cloud $\sim7 \times 10^6$~M\solar (Lis \& Goldsmith 1989), in agreement 
with
the general weakness of the \thco\ emission in Sgr~C compared to
Sgr~B2.


\subsubsection{The Morphology of Fe K-$\alpha$ and $^{13}$CO Line Emission}

Figure 5a shows contours of 
 $^{13}$CO line emission 
integrated over the velocity range between -77 and -22 \kms
superimposed on the grayscale distribution of 
Fe K-$\alpha$ 6.4 keV line emission from 
Sgr~C.   The strongest 
molecular line emission has a velocity of -65 \kms which identifies the 
Sgr~C molecular cloud (Liszt \& Spiker 1995). An isolated molecular 
feature near b$\sim-3'$ arises from the -120 to -90 \kms molecular cloud. The 
strongest 6.4 keV line emission G359.48--0.15 arises from the  
western portion  of the molecular cloud. 
 The second strongest 6.4 keV peak emission is about 2$'$ southwest 
of G359.48--0.15. These faint X-ray features are correlated better with the 
fainter lower velocity components near -40 \kms. It is clear that the 
overall distribution of the 6.4 keV emission follows the western edge of 
the -65~\kms molecular cloud, as there is no X-ray emission from the 
eastern 
edge of the molecular cloud. Figure 5b shows another version of Figure 5a 
except 
that contours of $^{13}$CO emission are superimposed on the grayscale 
X-ray continuum emission. 
 The continuum X-ray emission from the Sgr~C complex is 
also 
consistent with being distributed to the western edge of the -65 \kms 
molecular cloud. This suggests that the fluorescent 6.4 keV line emission 
and the X-ray continuum emission trace each other, though the 
line-to-continuum ratio 
may vary across the cloud. Moreover, the nonthermal filaments (see Figure1 \&3) 
are detected only at the western edge of the Sgr C cloud. 
These images suggest  the correspondence of K$\alpha$ line and nonthermal radio
emission at the western edge of the Sgr C molecular cloud.  

We also note that distribution of the equivalent width of the 
Fe~K-$\alpha$ line  is 
strongest  to the west of the -65 \kms molecular cloud 
with a value of 450 eV. This peak coincides with 
the peak of  $^{13}$CO 
emission at -65 \kms  at 
G359.47--0.15. Figure 5c shows contours of 
2-6 keV X-ray emission superimposed on the 6.4 keV equivalent width 
image of Sgr~C. The intensity of the Fe K$\alpha$ line emission toward 
Sgr~C is quite high, $\sim 6^{+2}_{-4} \times 10^{-6}$ ph cm$^{-2}$ 
s$^{-1}$ arcmin$^{-2}$, which is within at least a factor of $\sim$2 
larger than the average value in the central 20 pc of the Galactic 
center. Table 2 lists the parameters of the fit to the Sgr C region. 
The equivalent 
width of the Fe K-$\alpha$ line is not sensitive 
to the uncertainty introduced by the absorption. 


The distribution of 6.4 keV line emission based on ASCA observations was compared with that of CS (1-0) line emission 
from Sgr C (Murakami et al. 2001a).  However, the high negative velocity molecular gas with the velocity interval 
between -120 and -110 \kms was compared with their low spatial resolution X-ray data.  They assumed that the core of 
Sgr C is centered on the  high negative velocity feature  at -110 \kms. 
However, the core of 
Sgr C and the bulk of molecular gas is known to have a velocity of -65 \kms (see Figure 5a,b), as a number of ionized and 
molecular 
studies have indicated (e.g., Tsuboi et al. 1991; Liszt \& Spiker 1995; Lis \& Carlstrom 1994).  In order to make a 
more detailed comparison of our X-ray and radio continuum data with the -110 \kms molecular feature, we used the same 
CS (1-0) molecular line data (Tsuboi et al. 1999) that Murakami et al. (2001a) had used. 
The distribution of \thco\  is  quite
similar to CS(1-0) suggesting that 
both trace the overall molecular gas distribution in the region. 
Figures 6a, b show contours of 
CS(1-0) emission between -120 and -110 \kms velocity interval superimposed on the 6.4 keV line and 90cm continuum 
distributions, respectively. We agree with Murakami et al. assessment that K$\alpha$ line emission at the position of 
l=$359^0 26'$ lies at the edge of one of the CS peaks at this velocity interval. However. there is no evidence of 
strong K$\alpha$ emission counterpart to the CS peaks at positive galactic latitudes.  Figure 6b shows that the 
nonthermal Sgr C filament lies also at the edge of the -110 \kms velocity feature.  The morphology of a nonthermal 
filament lying at the edge of K$\alpha$-line emitting molecular gas is also made for the -110 \kms velocity feature of 
Sgr C.  Thus, the LECRs from the Sgr C filament could also be responsible for the origin of K$\alpha$ 6.4 keV line 
emission from the high negative velocity component, as discussed in more detail below. Thus, there are two issues that 
weaken the application of the irradiation model to the the -110 \kms cloud. One is that the \thco\ integrated intensity 
in the -200 to 200 km/s velocity interval averaged over the area mapped is 84 Kkm/s. The corresponding value for the 
-120 to -110 km/s velocity range is 9.5 Kkm/s. One of the questions that arises is that the X-ray radiation nebula 
model would thus have to explain why only 11\% of the molecular gas on the line of sight is irradiated assuming that 
the -110 and the -65 \kms velocity components are part of the Sgr C cloud.  The other is the lack of 6.4 keV emission 
from other peaks of the -110 \kms cloud facing toward Sgr A*.

There is no X-ray counterpart to the edge of the cloud in our Chandra as well  
as XMM data when compared with Figure 1 of Murakami et al. (2001a).
A  number of reasons could  be potentially responsible for this descrepancy. One is the
low-resolution map based on their  ASCA mesurements which could be contaminated by
point sources within 6.1-7 keV, the there are not enough point sources to account for the
excelss emission. The other is their linewidth being 
different,
6.1-7.1 versus 5.8 to 7.0 keV.  Lastly, time variability of the iron flux
is also a possibility over the course of 8 years.  However, the comparsion of the
iron flux from a knot-like feature at G359.48-0.15 with that of Murakami et al.
(2001a)
requires about 24 light years to
explain the shift in the centroid of this knot-like feature.
In addition, the derived EW (460+-100) eV seems to be lower
than the value (800 eV) obtained by Murakami et al. We believe 
the main reason for the difference is that we used different
assumptions for the background emission. Murakami et al. (2001a) used an
off-source region for the background, so their subtraction is both of
particle background and (an estimate of) the Galactic diffuse emission.
Our images were subtracted only by the particle background. Lastly, 
the excess emission could be due to a transient behind a cloud.


If the fractional \thco\ abundance in Sgr~C is lower than in Sgr~B2, the column 
density estimate could be higher. However, the $^{12}$C/$^{13}$C ratio is 
$\sim$20 in the Galactic center compared to the local value of 
77$\pm$7 (Wilson \& Rood 1994). With X(CO)=10$^{-4}$, the 
``nominal'' Galactic center \thco\ abundance is 
X($^{13}$CO)=4$\times10^{-6}$. The low value of X($^{13}$CO) = 
$10^{-6}$ which we used is already low and is more consistent 
with the local ISM value (e.g., Frerking et al. 1982). This 
suggests that the estimated \thco\ abundance can not be lower.
 
Another useful method that traces the column density of molecular
hydrogen is sub-millimeter emission.  Figure 7 shows a grayscale 850
$\mu$m image of Sgr C based on SCUBA observations (Pierce-Price et al.
2000) and contours of \thco\ line emission.  Two sub-millimeter peaks
are noted; the northeastern component is a site of star forming region
associated with the -65 \kms Sgr C cloud whereas the southeastern
component is unlikely to be related to the Sgr C cloud.  The peak
sub-millimeter emission which lies in the vicinity of the peak of the
molecular line emission has a flux density of 7.44 Jy per 15$''^2$.
The peak column density is then estimated to be $\sim6\times10^{23}$
cm$^{-2}$ using the value of 513 \msol Jy$^{-1}$ at 850 $\mu$m, as
derived by Pierce-Price et al.  (2000).  The mean value of the column
density of the Sgr C cloud from sub-millimeter observations is about
2$\times10^{23}$ cm$^{-2}$.  This value is consistent with earlier
estimates of the column density determined by Lis \& Carlstrom 1994).

\subsubsection{Spectral Index
Measurements}

We obtained a   spectral index  $\alpha$ (where flux density F$_{\nu} 
\propto 
\nu^{\alpha}$) 
distribution of Sgr C by using the 90 and 20~cm images
(Nord et 
al. 2004; Yusef-Zadeh, Hewitt, \& Cotton 2004), 
convolved with a FWHM Gaussian 12.6$''\times12.6''$. 
Radio continuum emission from Sgr C at 90~cm is an excellent probe 
of nonthermal sources  in this complex source. 
Figure 8a  shows a slice cut of  the spectral index distribution 
along the 
brightest  filament whereas Figure 8b shows the error plot of 
the spectral index distribution.  The spectral index is typically ranging between 
--1$\pm0.3$  and --0.7$\pm0.1$ to the south and  to the north of the 
linear filament  away from the Galactic plane, respectively. 
The emission from the Sgr C H\,{\sc ii} region at 20~cm  
is mainly responsible for the large  error in the value  of $\alpha$ to the south 
of the filament.  These  measurements are consistent with 
the spectral index values  between --0.7 and --1, as  determined between 20 and 6~cm 
(Liszt \& Spiker 
1995). 
We also compared the 3.6 and 2~cm images of the  bright nonthermal filament 
and found a steepening of the spectral index at shorter wavelengths. 

Figures 9a,b show the 3.6 and 2~cm images, respectively, with a resolution 
of 6.8$''\times6.4''$ (PA=$10.2^{\circ}$).  We note the vertical filament is 
detected prominently at 3.6~cm with a typical surface brightness of $\sim$ 3 
mJy beam$^{-1}$. However, the upper limit to the brightness of the filament 
is about 0.4 mJy beam$^{-1}$ at 2~cm. This filament was not detected at 1.2~cm either 
(Tsuboi et al. 1991). Accurate spectral index measurements 
restricting the {\it {uv}} range to 2 and 30 k$\lambda$ for both the 3.6 
and 2~cm data show that the spectral index of the bright filament is equal 
or steeper than -2.3$\pm0.5$. This suggests a break in the spectral index 
of nonthermal filaments from 90 to 2~cm, possibly due to the 
shorter lifetime of more relativistic particles at high energies.  Previous 
spectral index 
measurements of other Galactic center filaments show similar spectral index 
break  with  a value of $\sim-2$ at shorter wavelengths than at 
 6~cm (Lang, Morris \& Echevarria 
1999). The first detection of polarized emission from Sgr C, as shown in 
Figure 10,  is  described in the Appendix.  



\subsection{Galactic Center Molecular Clouds at $l>0^{\circ}$}
\subsubsection{6.4 keV Emission} 

With the exception of the Sgr C molecular cloud 
 that lies beyond Sgr A and at the
negative longitude side of the Galactic center, the largest
concentration of dense and warm molecular clouds is known to be
distributed near Sgr A and  in the positive longitude side of the Galactic center
region (e.g., Morris \& Serabyn 1996; Tsuboi, Handa \& Ukita 1999).
Four   of the well-known molecular cloud complexes 
are  ``20 \kms cloud'' (M--0.13--0.05) distributed near Sgr A, 
the ``45 \kms cloud'',
the ``-30 \kms cloud'', both of which extend between Sgr A and the
radio arc at l$\sim0.2^{\circ}$, and the molecular ridge extending from
molecular cloud M0.25--0.01 to Sgr B2.  The 45 and -30 \kms clouds lie
on one side of the radio arc and overlap with each other spatially.
The largest concentration of bright and long nonthermal radio
filaments are also found on the positive longitude side of the
Galactic center (Yusef-Zadeh, Morris, \& Chance 1984).  The radio arc
and its adjacent molecular clouds can potentially provide evidence for
a mixture of interacting thermal and nonthermal features.  Given that
relativistic particles of of the Sgr C filaments are interacting with
its molecular cloud, we believe that the interaction of nonthermal
particles of the radio arc with its adjacent molecular clouds can also
explain the nature of 6.4 keV emission, as described below.

To illustrate the proposed physical association between 
nonthermal radio filaments, the 45 and -30 \kms molecular clouds and the 6.4 keV line emission, 
Figures 11a,b show grayscale views of the non-thermal radio filaments of the radio arc and Sgr A 
at  20~cm  against the fluorescent 6.4 keV image and ammonia distribution (G\"usten et al. 1981; 
Yusef-Zadeh 1986).  Figure 11a shows the distribution 
of the K$\alpha$ EW line emission in green color and a 20~cm continuum image in red color whereas the 
NH$_3$(1,1)  line emission from the ``20 and 45 \kms clouds'' 
are displayed  in Figure 11b as 
contours 
with a spatial 
resolution 
of 40$''$.  Given that 
ammonia line observations 
have lower resolution than radio continuum and X-ray observations, the overall distribution of bright 
knots of K$\alpha$ 
EW line emission correlate well with peaks of ammonia line emission associated with the ``45'' 
\kms 
molecular cloud.   We note that there is no evidence for strong compact or diffuse K$\alpha$ line 
emission 
from the  ``20 \kms cloud''. This cloud is one of the most prominent 
dust continuum features in the Galactic center showing high mass and column density. 
Furthermore,  nonthermal filaments have been detected toward this cloud. 
Similarly, another massive cloud 
that shows a  lack of EW of K$\alpha$ line emission is M+0.25+0.01. 
Individual 
compact 
ammonia clumps of the 45 \kms molecular cloud that 
have 6.4 keV line correspondence, as shown in 
Figure 11b, are M0.06--004, M0.10--0.01, M0.11--0.08, M--0.02--0.07 and 
M0.07--0.08 (e.g., Serabyn \& 
G\"usten 1987; Lang, Goss, \& Morris 2002). The  molecular clump 
M0.11--0.08 lies closest in projection  to the nonthermal filaments of the arc 
has been extensively  mapped recently in CS, SiO and H\thco$^+$ lines. This 
clump has  velocities ranging between 15 and 45 \kms  and has 
 a strong  6.4 keV line counterpart 
(Tsuboi, Ukita, \& Handa 1997; Yusef-Zadeh, 
Law, \& Wardle 2002; Handa et al. 2006).

Another cloud that is dynamically coupled to the thermal component of
radio arc is the -30 \kms cloud (Serabyn \& G\"usten 1987).  This
cloud is distributed on the positive latitude side of the arc and is
associated with thermal gas of the arched filaments which is ionized
by the Arches cluster G0.121+0.017 (e.g., Cotera et al.  1996; Figer
et al.  1999; Lang, Goss \& Morris 2002).  The comparison of the
distributions of the CS molecular gas and K${\alpha}$ line emission
suggests that the K$\alpha$ line peaks follow the distribution of the
CS emission (Serabyn \& G\"usten 1987).  In fact, one of the ammonia
peaks M--0.02--0.05 (--15 \kms), as shown in Figure 11b, coincides with a
southern extent of the -30 \kms molecular cloud.  M--0.02--0.05, as well
as the CS peaks 1 and 2 identified by Serabyn \& G\"usten (1987) have
6.4 keV line correspondence.  Lastly, another knot-like K$\alpha$ line
emission G0.12+0.00 distributed in the immediate vicinity of Arches
cluster G0.121+0.017 appears to be associated with a CS molecular
emission (Lang, Good \& Morris 2002; Serabyn \& G\"usten 1987).

The hook-shaped structure noted in the the K$\alpha$ EW line
distribution correlates extremely well with the ammonia distribution,
as noted in Figures 11a,b.  In addition to nonthermal emission from
magnetized filaments and the  recent detection of nonthermal emission from the
Arches cluster provide another source for the interaction of cosmic
rays with molecular gas to explain the origin of the 6.4 keV line
emission.  Although the total mass of molecular gas and radio
continuum flux of nonthermal emission in the arc region are much
higher than that those in Sgr C, the column density of molecular gas
and the EW of K$\alpha$ line of about few hundred eV are similar to
those of Sgr C. With the exception of G0.11--0.08 molecular cloud which
is closest to the filaments in projection having a hydrogen column
density of 7-8$\times10^{23}$ cm$^{-2}$, the peaks of the 45 \kms and
-30 \kms clouds show column densities of of few times 10$^{23}$
cm$^{-2}$ (G\"usten et al.  1981; Serabyn \& G\"usten 1997).  The
density and temperature of molecular gas from these clouds are also
estimated to be about few $\times10^4$ cm$^{-3}$ and 50-80 K,
respectively.

In the context of low-energy cosmic ray model, these characteristics
imply that cosmic ray particles penetrating molecular clouds should be
uniform with similar metallicity as discussed in more detail in $\S$4.2
We measured the K$\alpha$ line and continuum X-ray flux of the arc by
taking a spectrum over an area of 123 arcmin$^2$ encompassing both 45
and -30 \kms molecular clouds.  An absorbed power-law plus Gaussians
were used to model all of the line emission.  This model is not
physically motivated, but allows to measure fluxes and equivalent
widths.  The de-reddened continuum flux between 2-8 keV, Fe K$\alpha$
EW and Fe K$\alpha$ flux are 5.6$\times10^{-11}$ erg cm$^{-2}$
s$^{-1}$, 670 eV and 3.5$\times10^{-4}$ ph cm$^{-2}$ s$^{-1}$,
respectively.  The parameters of the fits to Sgr C, the radio arc, Sgr
B1 and Sgr B2 are shown in Table 2.  These parameters are 
derived\, assuming a model of a 
power-law continuum under which the
absorption columns are estimated to be lower than in\, thermal
plasma models by a factor of 2--3.  This results in a 30\% lower estimated
flux for  the line from Sgr C, which should be taken as the systematic
uncertainty in the measurement.
\clearpage
\begin{deluxetable}{lrrrrrr}
\tablecolumns{7}
\tablewidth{0pc}
\tablecaption{Parameters of X-ray Fit to Galactic Center Clouds}
\tablehead{
   \colhead{Region} &
   \colhead{Area}   &
   \colhead{Fe K$\alpha$ Flux} &
   \colhead{Fe K$\alpha$ EW}  &
   \colhead{Flux (2--8 keV)}   &
   \colhead{De-reddened Flux} \\
   \colhead{} &
   \colhead{(arcmin$^2$)} &
   \colhead{(ph cm$^{-2}$ s${^-1}$)} &
   \colhead{(eV)} &
   \colhead{(erg cm$^{-2}$ s$^{-1}$)} &
   \colhead{(erg cm$^{-2}$ s$^{-1}$)} \\}
\startdata
Sgr C     & 77  &  $7.2(0.2)\times10^{-5}$ & 470(100) &    $9.1(0.1)\times10^{-12}$ & $1.2(0.2)\times10^{-11}$ \\
Radio Arc & 123 &  $3.5(0.1)\times10^{-4}$ &  670(50) &  $3.7(0.1)\times10^{-11}$ & $5.6(0.1)\times10^{-11}$    \\
Arches Cluster & 1.6 & $1.3(0.1)\times10^{-5}$ & 810(200) & $9.1(0.1)\times10^{-13}$ & $1.2(0.2)\times10^{-12}$ \\
Sgr B1  & 77  &  $8.6(4)\times10^{-5}$ &  570(70) &  $7.6(0.1)\times10^{-12}$ & $9.7(0.7)\times10^{-12}$  \\
Sgr B2  & 96  &  $2.4(0.1)\times10^{-4}$ & 1150(150) &  $9.5(0.1)\times10^{-12}$ & $1.1(0.1)\times10^{-11}$   \\
\enddata
\end{deluxetable}
\clearpage
To examine the impact of low-energy comic ray particles with molecular
gas throughout the Galactic center, we also list in Table 2 the fitted
parameters to two other Galactic center molecular cloud complexes Sgr
B1 and B2.  These clouds are associated with the molecular ridge that
extends between G0.25+0.01 and Sgr B2, as shown in Figure 2.  Both
these clouds have been extensively studied in millimeter and
submillimeter wavelengths, as well as in K$\alpha$ line (e.g.,
H\"uttemeister et al.  1993a; Lis \& Carlstrom 1994; Murakami et a.
2001).  The regions from which spectra were extracted from Sgr C, the
radio arc, Sgr B1 and Sgr B2 are drawn on a large-scale 6.4 keV EW
line distribution, as displayed in Figure 2.  A more detailed study of
the individual 6.4 keV line features will be given elsewhere.

\subsubsection{TeV  Emission}

Recent observation with HESS has discovered large-scale diffuse TeV
emission from the inner 200 pc of the Galaxy (Aharonian et al.  2006).
The morphology of diffuse emission correlates well with the
distribution of CS molecular clouds, thus suggesting that the
$\gamma$-ray emission is a product of the interaction of cosmic rays
with interstellar gas near the Galactic center.  These authors show
that the spectrum of TeV emission from resolved clouds toward the
Galactic center has a a photon index $\Gamma\sim 2.3$ which is harder
than in that in the Galactic disk.  They note that the $\gamma$-ray
flux above 1 TeV is a factor of 3-9 times higher than in the Galactic
disk and argue for an additional population of cosmic rays in this
unique region.  They propose that the TeV emission is due to hadronic
interaction of cosmic rays with the target material.  Their argument
against the importance of TeV electrons is the short lifetime of
their energies.

Given that the target material is the same, we compared the
distribution of TeV emission with that of the 6.4 keV emission from
molecular gas toward the Galactic center.  Figure 12a shows contours of
diffuse TeV emission superimposed on a grayscale image of K$\alpha$
6.4 keV line emission.  The prominent TeV peaks are seen toward Sgr C,
the radio arc and Sgr B. Due to the subtraction of the compact and
bright source coincident with Sgr A*, it is not clear if the weak
feature near G359.83--0.1 is real or due to an artifact of subtraction
(J. Hinton, private communication).  As expected, there is a strong
concentration of 6.4 keV knots which correlates well with the peaks of
TeV emission toward Sgr C, Sgr B and the arc.  The brightest TeV
emission from diffuse galactic center molecular clouds is detected
toward the radio arc followed by Sgr B2 and Sgr C. We also   note  that
both the TeV  and K$\alpha$ line emissions from the well-known ``20 \kms'' molecular cloud 
G--0.13--0.08 are weak.  To illustrate the relationship between  molecular gas, nonthermal filaments  and 
the fluorescent line emission, 
 Figure 12b shows the contour distribution of submillimeter emission distributed 
against a  20~cm  continuum emission (in green) and 6.4 keV emission (in red). All the prominent 
molecular clouds are represented in the submillimeter image.  

\section{Discussion}

There are three different types of observations that are used in support of 
large-scale high energy activity in the inner few hundred pcs the Galactic 
center. These observations suggest that this activity is  more 
pronounced than that observed in the disk of the Galaxy for the following 
reasons. One is the 
distribution of nonthermal radio emission traced by 
magnetized radio filaments, a 1$^{\circ}$   scale 
diffuse  structure, known as the Galactic center lobe, a 
high 
concentration of confirmed supernova 
remnants such as G0.0+0.0 or Sgr A East (Ekers et al. 1983), G0.3+0.0 (Kassim \& Frail 1996), G0.9+0.1 (Helfand \& 
Becker 
1987), G1.0-0.1 or Sgr D (Downes et al. 1979) and G359.1  
-0.5 (Downes et al. 1979) as well as  nonthermal 
emission from colliding winds of massive 
stars in dense stellar clusters. These sources, altogether
signify regions of enhanced synchrotron emissivity  (e.g., Gray 
1994;  Nord et 
al. 2004; Yusef-Zadeh et al. 2004; Law 2006). 
The recent detection of diffuse 
low-frequency radio continuum 
emission (LaRosa et al. 2006), as well as other indirect measurements 
(Yusef-Zadeh 2003) suggest that the large-scale diffuse distribution of the 
magnetic field can not be too different than 
that observed 
in the Galactic disk.  This is consistent with the picture that the Galactic center  nonthermal radio 
filaments are 
expected to 
have a stronger magnetic field  than that of the surrounding diffuse and weakly magnetized medium. 
Recent interpretation of the nonthermal filaments suggests that  kinematic dynamo enhances the magnetic field 
in a medium which is initially threaded by weak magnetic field (Bodyrev \& Yusef-Zadeh 2006).  
 Thus, we suggest 
that 
cosmic ray electron density of the diffuse gas must be enhanced in the Galactic center  region in order to account
for the excess  synchrotron emissivity detected toward the central 200 
pcs of the 
Galaxy.
Another recent measurements  come from H$_3^+$ and H$_3$O$^+$ 
observations  suggesting that 
the ionization rate 
due to cosmic rays has to be higher by one to two  orders of  
magnitude in the Galactic center region than in the disk (Oka et al. 
2005; van der Tak et al. 2006). 
Lastly, the recent discovery of diffuse TeV emission from the Galactic 
center molecular clouds shows enhanced $\gamma$-ray emission at higher 
energies having a spectrum that differs from that of the cosmic rays 
 in the disk of 
the Galaxy.

\clearpage
\begin{deluxetable}{lrrrrrr}
\tablecolumns{7}
\tablewidth{0pc}
\tablecaption{Parameters of the LECR Model}
\tablehead{
   \colhead{Region} &
   \colhead{Area}   &
   \colhead{90~cm  Flux } &
   \colhead{Energy Density}  &
   \colhead{Magnetic Field}   &
   \colhead{spectral index } \\
   \colhead{} &
   \colhead{(arcmin$^2$)} &
   \colhead{(Jy)} &
   \colhead{(eV cm$^{-3}$)} &
   \colhead{(mG)} &
   \colhead{($\alpha$)} \\}
\startdata
Sgr C          & 0.024  &  0.05  & 1200 &    0.22& --0.7 \\
Radio Arc      &    285 &  260   &  19  &   0.029& 0.0\\
Arches Cluster & 1.6    & 0.36   & $6\times10^4$  &  1.6   & --1 \\
Sgr B1         & 77     &3.2     &  23  &  0.03  & --0.7\\
Sgr B2         & 96     & 18     &  51  &  0.045 & --0.7 \\
\enddata
\end{deluxetable}
\clearpage
Based on the above arguments, it is natural to consider the possibility 
that enhanced 
cosmic rays in the Galactic center region could play an important role in 
accounting for much of the observed high energy activity in this region. 
This hypothesis stem from the fact that a great deal of warm and dense 
molecular gas is  distributed in the central region (e.g., Rodriguez-Fernandez et al. 2001), 
thus, the 
widely distributed molecular 
gas can be used as an excellent target by the impact of 
enhanced cosmic  rays.   In 
particular, the origin of the observed 6.4 keV emission throughout the 
Galactic center region can be tied to the impact of low-energy 
cosmic 
ray electrons with molecular gas. The role of low-energy cosmic rays 
was  first  proposed 
to explain the origin of the Galactic X-ray ridge (Valinia et al. 2002). This 
model was subsequently   
applied to G0.11--0.08 (Yusef-Zadeh, Law, \& Wardle 2002). We argue 
below that this model can also be applied to Sgr C, as well as 
other  Galactic center molecular clouds. In addition, we argue an 
alternative model to explain the origin of TeV emission from molecular 
clouds near the Galactic center. We suggest that the high-energy component 
of electrons   can potentially be responsible to inverse Compton scatter 
the intense submillimeter emission to account for the  origin of 
$\gamma$-ray emission. 

\subsection{The Irradiation versus LECRs Model}

One of the main results of our analysis of the millimeter and X-ray
data is that the distributions of the 6.4 keV line and 2-6 keV
continuum emission arise from the western edge of the $^{13}$CO
molecular cloud.  The main body of the Sgr~C cloud does not show any
evidence of X-ray emission except from its western edge which faces
away from the direction of Sgr~A*. The origin of fluorescent line
emission resulting from irradiation by a hypothetical transient source
associated with Sgr~A* has been applied to the Galactic center
molecular clouds (e.g., Sunyaev, Markovitch, \& Pavlinsky 1993; Koyama
et al.  1996; Murakami et al.  2001a; Park et al.  2004).  However,
unlike Sgr B2, the X-ray emission from Sgr C faces away from the
suggested illuminating source, Sgr A*.  The column density
of molecular hydrogen  10$^{23}$ -- 10$^{24}$
correspond to Thompson optical depth $\tau_T$ of 0.1 to 1
(Sunyaev \& Churazov 1998).  If the illuminating source is outside
the cloud, the expected equivalent width (EW) exceeds 1 keV for an
optically thin or thick case.  The combined distribution of 6.4 keV
line emission from Sgr C with respect to the bulk of molecular gas as
well as Sgr~A* when combined with the estimated EW of $\sim$ 470 eV
toward Sgr C suggest strongly that the irradiation model is unlikely
to be applicable to Sgr~C.

We propose an alternative model to explain the origin of X-ray
emission from Sgr~C in terms of the impact of low energy cosmic rays
(LECRs) with molecular gas.  The motivation for such a process stems
from the fact that Sgr~C is surrounded by a large concentration of
nonthermal radio emission associated with radio filaments.  Given that
the centroids of X-ray and radio continuum features are displaced with
respect to each other, it is remarkable that all the X-ray continuum
features, labeled in Figure 1a, appear to be located in the vicinity
of peak nonthermal radio continuum emitting features associated with
the Sgr C nonthermal filaments.  Furthermore, the 6.4 keV line
emission generally follows the X-ray continuum emission for almost all the X-ray
features with the exception of G359.32--0.16 which could be a pure 
nonthermal continuum source 
with no X-ray emission lines or molecular counterpart. 
The  model proposed here can naturally explain  the origin of 
hot molecular gas observed throughout the Galactic center molecular clouds. 

In the context of LECR model, the remarkable mixture of thermal and
nonthermal radio components in Sgr~C can produce not only the 6.4 keV
line emission but also the nonthermal bremsstrahlung emission from
Sgr~C. As pointed out in $\S$2, the X-ray spectrum could be
equally fitted by a combined soft thermal and a power-law spectrum as
long as the metallicity is higher than solar value.  The power-law
spectrum of the continuum emission from Sgr~C is consistent with the
LECR model, as described quantitatively below.

What about nonthermal filaments  that have X-ray counterparts
with no 6.4 keV counterpart?  We believe that a  flat spectrum of these filaments  
reduces the number of low energy cosmic rays for  interaction  with molecular gas but
increases the number of high energy particles needed to  produce synchrotron X-ray
emission.  The nonthermal X-ray filament discovered by Sakano et al. (2003) coincides 
exactly with a flat nonthermal radio source which  is thought to be associated with 
``the 20 \kms molecular 
cloud'' ( Ho et al.  1985; Coil \& Ho 2000; Yusef-Zadeh et al. 2005). Assuming that this cloud is
near the Galactic center, the lack of 
strong 6.4 keV
and TeV emission from this molecular cloud could be explained by the flat spectrum of 
its   nonthermal filament.

The source of the enhanced cosmic-ray electrons is acceleration at the
interaction site between the Sgr~C molecular cloud and the Sgr~C
nonthermal radio filaments.  The distribution of diffuse and
filamentary X-ray gas lying at the edge of the filaments imply the
presence of relativistic electrons in the vicinity of the diffuse
X-ray source.  However, the X-ray and radio features, as identified in
Figures 1a and 3, are either displaced from each other or the X-ray
features peak at the edge of nonthermal radio filaments.  A
displacement was also observed between the peak emission from radio
filaments of the continuum arc and the M0.11--0.08 molecular cloud.
The filaments of the arc and G0.11--0.08 are thought to be interacting
with each other (Tsuboi et al.  1997; Oka et al.  2001).  In addition,
LECR model has been successfully applied to this cloud (Yusef-Zadeh,
Law, \& Wardle 2002).  So, the displacement of radio emitting
filaments and molecular cloud may be widespread in clouds that are
interacting with nonthermal radio filaments.  We believe this
characteristic can be explained by the geometry of the interacting
non-thermal filaments with respect to the distribution of the
molecular cloud.  As described below, the low-energy 
relativistic particles interacting with neutral gas lose energy much
more than the GeV particles that produce radio emission at high
frequencies.  The high-energy particles could be oblivious to the
presence of molecular gas in their vicinity.  

Interaction of cosmic rays with molecular gas has recently been argued
to explain the origin of TeV emission from diffuse molecular gas.  The
grayscale image in Figure 12 shows contours of TeV emission from the
Galactic center region superimposed on a large-scale distribution of
Fe K$\alpha$ line EW emission.  There appears to be a correlation
between the 6.4 keV EW line emission and TeV emission.  The strongest
TeV emission arises from the region near the arc which is coincident
with an EGRET source, as well as a soft $\gamma$-ray source 1743.1-2843
detected with INTEGRAL between 50 and 100 keV (Belanger et al.  2006).
This source has been suggested to be produced as a result of the
impact of GeV particles of the nonthermal filaments and G0.11--0.08
(Yusef-Zadeh, Law, \& Wardle 2002).  Whatever the mechanism
responsible for the $\gamma$-ray emission from this source, it is
suggestive that cosmic rays interacting with dense Galactic center
molecular clouds are responsible for both types of X-ray, as well as
soft and hard $\gamma$-ray emission.  Apart from supernova remnants
and nonthermal filaments, massive stellar clusters can also provide
relativistic particles generated as a result of wind-wind collision in
a dense stellar cluster.  The Arches cluster is one example that has
shown a flux of $\sim$100 mJy of nonthermal emission from the cluster
(Yusef-Zadeh et al.  2003).


\subsection{Injection of Cosmic Rays into Molecular Clouds}

We now calculate production rate of Fe K$\alpha$ photons associated
with the injection of $dn(E)/dt$ electrons per unit energy interval
per unit time into a cloud with column density $N_\mathrm{H}$.
The column density of the Sgr~C cloud in the vicinity of the X-ray
emission is $N_\mathrm{H}\sim 10^{23}$\,cm$^{-2}$. 
This is sufficient
to stop electrons with energies below $\sim 1$\,MeV; the rapid
increase in penetrating power with energy means that higher energies
pass through this column with little fractional energy loss (see, e.g.
ICRU 1984).  
Electrons with initial energy below a critical value
$E_c(N_\mathrm{H})$ (1\,MeV for $10^{23}$\,cm$^{-2}$) are stopped in
the cloud.  A single electron injected with energy $E<E_c$ produces
\begin{equation}
    z \omega_K f_\alpha \int_0^E 
    \sigma_K(E')\,\frac{-dN_s}{dE'}\,\,dE'
    \label{eq:single_e}
\end{equation}
Fe K$\alpha$ photons while coming to rest, where $z$ is the abundance
of iron relative to hydrogen ($z_\odot=2.8\times10^{-5}$), $\omega_K =
0.342$ is the fraction of K-shell ionizations that produce
K-characteristic X-rays, $f_\alpha = 0.822$ is the fraction of those
that produce a K$\alpha$ photon, and $N_s(E)$ is the column density
traversed by electrons of initial energy $E$ before coming to rest.
In steady state the net K$\alpha$ production rate produced by the 
injected electrons with energies $E<E_c$ is therefore
\begin{equation}
    Q_1 = z \omega_K f_\alpha \int_0^{E_c} \frac{dn(E)}{dt}\,\int_0^E 
    \sigma_K(E')\,\frac{-dN_s}{dE'}\,\,dE'\,\,dE\,.
    \label{eq:multiple_e}
\end{equation}
Reversing the order of integration yields
\begin{equation}
    Q_1 = z \omega_K f_\alpha \int_0^{E_c} 
    \sigma_K(E)\,\frac{-dN_s}{dE}\,\frac{dn}{dt}(E<E'<E_c)\,\,dE\,.
    \label{eq:Q_thick}
\end{equation}


Electrons that enter the cloud at higher energies only lose a 
fraction of their energy; for simplicity we neglect this energy loss 
in calculating their contribution to the Fe K$\alpha$ production 
rate, and write:
\begin{equation}
    Q_2 =  z \omega_K f_\alpha N_\mathrm{H}\,
    \int_{E_c}^\infty \frac{dn}{dt}(E)\,
    \sigma_K(E)\, dE \,.
    \label{eq:Q_thin}
\end{equation}
This ``thin-target'' approximation is adequate for our purposes here
because $\sigma_K$ varies
by only a factor of two between 10\,keV (e.g.\ Tatischeff 2003) and 1\,GeV.

For $\sigma_K(E)$ we use the semi-empirical expression of Quarles
(1976) with an ionization threshold energy $I=7.1$\,keV appropriate
for neutral or almost-neutral iron.  The spectrum of injected
electrons is assumed to be a power-law ($E^{-p}$) between 10\,keV and
1\,GeV, normalized so that the corresponding power is 1\, erg/s.
The corresponding production rate $q=Q_1+Q_2$ of Fe K$\alpha$ photons per erg of
injected electron energy is calculated using eqs (\ref{eq:Q_thick})
and (\ref{eq:Q_thin}) as a function of cloud column density.

The results are plotted in Figure 13a 
for electron spectral indices $p$
ranging from 2 to 3 and a solar iron abundance.  For a given particle
spectral index $p$, the efficiency of $K\alpha$ production increases
with column density, eventually flattening when the column is
sufficient to stop the bulk of the injected electrons within the
cloud.  For the $p=2$ injection spectrum, which has equal energy per
decade between 10\,keV and 1\,GeV, this occurs above $10^{25}\ut cm -2
$, which can only stop electrons with energies below 0.1\,GeV. Steeper
spectra have particle energies increasingly concentrated towards
10\,keV and so the flattening of the K$\alpha$ production rate occurs
at successively lower column densities.  These results are consistent
with those of Tatischeff (2003) who found $q\sim 50 \u ph \ut erg -1 $
for 10--100\,keV electrons with an $E^{-2}$ spectrum.
We conclude from Figure 13a that 
typically $q\sim 50 \, z/z_\odot \u ph \ut erg -1 $ for typical spectral indices 
and cloud column densities, increasing to $\sim 100\, z/z_\odot$ for hard 
electron spectra and high column densities. 
The bremstrahlung emission at 6.4 keV was estimated similarly; the resulting 
equivalent width is plotted in Fig. 13b.   For
$z = z_\odot$, the equivalent width  varies between 250 and 300 eV for particle 
spectral index $p$ varying between 3 and 2.
There is little dependence on column except for $p$ close to 2 and columns in 
excess of $10^{24} \ut cm -2 $, where the
equivalent width increases from 300 to 340 eV as the column increases towards 
$10^{25} \ut cm -2 $. The Galactic center molecular clouds have typical 
metallicity that is on the average twice  the solar value (e.g., 
Giveon et al. 2002; Rudolph et al. 2006). Thus,  the production of K$\alpha$ 
line emission per erg of energy and the EW of K$\alpha$ emission, as 
  shown in Figure 13,  should be increased by a factor of two when applied to the 
Galactic center molecular clouds.

\subsection{Application of Low Energy Cosmic Ray Model}
 
To estimate the flux of electrons into the cloud, we assume that they
diffuse from their acceleration site to the cloud edge, and then
freely stream into the cloud because ion-neutral damping suppresses
the magnetic fluctuations responsible for the diffusion.  If the
energy density of the electrons external to the cloud is $U$, then the
flux into the cloud is $\sim cU/4$, giving an injection rate $cUA/4$
where $A\sim \Omega d^2$ is the cloud surface area impinged upon by
electrons, $\Omega$ is the solid angle of the X-ray emission and $d$
is the distance to the cloud from the Earth.  The production rate of
Fe K$\alpha$ photons is $Q\approx cU\Omega d^2q/4$.  The intensity of
Fe K${\alpha}$ photons at the Earth is then \begin{equation}
I_{\mathrm{K}_\alpha} = \frac{Q}{4\pi d^2 \Omega} = \frac{cUq}{16\pi}
\approx 8\times10^{-6}
\left(\frac{U}{10^3\,\mathrm{eV\,cm^{-3}}}\right)
\left(\frac{z}{2z_\odot}\right)
\mathrm{\,ph\,s^{-1}\,cm^{-2}\,arcmin^{-2}}
    \label{eq:IKalpha} 
\end{equation} 
comparable to the observed peak Fe K$\alpha$ intensities if U$\sim
1000$ eV\,cm$^{-3}$.

\def\fra #1 #2 {{\textstyle \frac{#1}{#2}}}
\def\ee #1 {\times 10^{#1}}
\def\ut #1 #2 { \, \mathrm{#1}^{#2}}
\def\u #1 { \, \mathrm{#1}}
\def\kms {$\,\mathrm{km\,s}^{-1}$}
\def\persec {\, \hbox{s}^{-1}}
\def\percc {\,\mathrm{cm}^{-3}}
\def\NOH{N_{\mathrm{OH}}}
\def\la{\lower.4ex\hbox{$\;\buildrel <\over{\scriptstyle\sim}\;$}}
\def\ga{\lower.4ex\hbox{$\;\buildrel >\over{\scriptstyle\sim}\;$}}
\def\refitem{\par\noindent\hangindent\parindent}


We apply this model to some of the sources in the Galactic centre
region.  In each case we assume that the nonthermal radio emission is
produced by a power-law spectrum $n(E)\propto E^{-p}$ of electron
energies between 10\,keV and 1\,GeV, assume that the depth of the
emitting region is of the order of its diameter and calculate the
energy density of cosmic-ray electrons assuming that they are in
equipartition with the magnetic field.  Eq.  (\ref{eq:IKalpha}) is
then used to compute the predicted Fe K$\alpha$ flux; when the source
of particles is assumed to be either embedded within the cloud or
completely surrounds it this estimate is increased by a factor of 4 to
allow for the increase in exposed surface area exposed to the incident
cosmic rays (ie.  $4\pi r^2 $ rather than $\pi r^2$).  Note that
although the incident electron spectrum may deviate from a power law
below 100 keV because of ionization losses, this is not a major
uncertainty as these electrons contribute $\la 20$\% of the K$_\alpha$
flux.  The model parameters needed to match the observed quantities
for each source are listed in Table 3.

\subsubsection{Sgr C}

The brightest filament associated with Sgr C has a typical synchrotron
flux at 90~cm of 50 mJy per $12.6''\times 6.8''$ beam with a
$\nu^{-0.7}$ spectrum.  This is a lower limit to the
 flux of
synchrotron emission because diffuse synchrotron emission, free-free
absorption by intervening thermal gas, as well as contribution from
weak nonthermal filaments have not been accounted for.  The
equipartition field is 0.22\,mG with an electron energy density of
$1200\u eV \ut cm -3 $.  This implies a peak K$\alpha$ flux of $9.7\ee
-6 \u ph \ut cm -2 \ut arcmin -2 $ which is about 1.5 times that
observed.  We have used a particle spectral index $p=2.4$ (where
$n(E)\propto E^{-p}$) and equipartition between electrons and magnetic
field.  The measured spectral index values between 90 and 20~cm
wavelengths agree well with the energy density in relativistic
electrons required to match the synchrotron emission.

\subsubsection{The 45 and -30 \kms clouds near the Radio Arc}

The filaments of the radio arc has a flux of $\sim$28\,Jy at 90~cm over
its $3'\times20'$ area, with a $\nu^{-0.5}$ spectrum.  This spectrum
is consistent with that measured between 408 and 160 MHz (Yusef-Zadeh
et al.  1986).  Most of the diffuse nonthermal emission is likely to
be resolved out so we used the Green Bank observations of this region
(Law et al.  2006) and measured a total integrated flux of 200 Jy over
285$'^2$.  This region includes the linear filaments of radio arc as
well as other weaker nonthermal filaments that have recently been
discovered (Reich 2003; Yusef-Zadeh et al.  2004; Nord et al.  2004)
within the  area of the extracted X-ray spectrum (Figure 11 and the box area 
in Figure 2).  Due to
the contamination by thermal H\,{\sc ii} features, as well as the background
emission at 90~cm, the flux of nonthermal emission has a large
uncertainty.  Using the recent 20~cm continuum image which includes the
zero spacing flux (Yusef-Zadeh et al.  2004), we found a similar flux
of about 200 Jy which is consistent with the 90~cm measurements having
a flat spectral index between 20 and 90~cm (Yusef-Zadeh et al.  1986).

Adopting 28\,Jy of flux density of the nonthermal filaments of the arc
 at 90\,cm and assuming that the arc 
consists
of filaments of thickness $30''$, then the equipartition field is
0.07\,mG and the energy density of electrons is $110 \u eV \ut cm -3 $
The predicted K$\alpha$ flux, $4\ee -5 \u ph \ut cm -2 \ut s -1 $, is
much lower than that measured ($3.5\ee -4 \u ph \ut cm -2 \ut s -1 $).
If we assume instead that the source is embedded in a diffuse
distribution of low-energy cosmic-ray electrons over 285 arcmin$^2$
with a source depth of 15$'$, then we find that an energy density of
38 $\u eV \ut cm -3 $ is required, with field strength 0.039\,mG and
synchrotron flux of 860\,Jy.  If the column density of the clouds is
$\sim 10^{24} \ut cm -2 $ and $p\sim 2$ then $q$ increases to $100
z/z_\odot \u ph \ut erg -1 $; the required energy density, field
strength and diffuse synchrotron flux are reduced to 19 $\u eV \ut cm
-3 $, 0.0289\,mG and 260\,Jy respectively.

\subsubsection{The Arches Cluster}

As discussed $\S$3.2, we found a knot-like K$\alpha$ emission in the
vicinity of the arches cluster.  The source of cosmic ray particles
injecting into the cloud surrounding the Arches cluster can either be
the nonthermal filaments of the arc or nonthermal particles generated
by cluster.  Figure 14 shows a detailed view of the arches cluster
which coincides with a compact nonthermal radio continuum source (in
dark) at 90~cm (Yusef-Zadeh et al.  2003) against the grayscale
distribution of the EW of 6.4 keV line emission.  An X-ray spectrum
was extracted from the region shown in Figure 14 which roughly
encloses the 600 eV contours.  The extracted region, as shown in Figure 2, 
is  defined by an
ellipse centered on $l$=7$' 33.7''$, $b$=$1' 08.6''$ having a major and
minor axis of 62.7$''$ and 29.1$''$ (PA=25$^{\circ}$), respectively.  The
parameters of the iron line emission are shown in Table 2 having Fe
K$\alpha$ EW of 810 eV. The peak EW emission in Figure 14 is as much
as 1000 eV.

The nonthermal 327\,MHz flux from the arches cluster is 91\,mJy over a
7$''$ by 13.5$''$ region (Yusef-Zadeh et al.  2002), with a spectral
index $\alpha\approx 1$ (for $S\nu\propto\nu^{-\alpha}$).  The
equipartition magnetic field is 1.1\,mG and the energy density of
electrons is $3\ee 4 \u eV \ut cm -3 $.  The predicted K$\alpha$ flux,
$6.5\ee -6 \u ph \ut cm -2 \ut s -1 $, agrees with that observed by Wang
et al.\ (2006), i.e.\ $6.2\ee -6 \u ph \ut cm -2 \ut s -1 $.  However,
if we use the integrated K$\alpha$ flux from the region shown in
Figure 14 and Table 2, the required 90~cm flux increases to 360 mJy of
nonthermal emission from the arches cluster with an equipartition
magnetic field of 1.6 mG. The additional flux of nonthermal emission
must arise from the diffuse nonthermal emission surrounding the
cluster or the emission from the nonthermal filaments of the arc.

\subsubsection{Sgr B2 and Sgr B1}

Radio continuum observations of the Sgr B complex shows two major
components Sgr B1 (G0.5--0.0) and Sgr B2 (G0.7--0.0), as well as an
intervening source G0.6--0.0 in the middle of these extended H\,{\sc ii}
regions.  (e.g., Mehringer et al.  1992).  Diffuse thermal emission
dominates Sgr B1 whereas numerous ultracompact H\,{\sc ii} regions overwhelm
the emission in Sgr B2.

The continuum emission from Sgr B2 is mainly due to several dozen
ultracompact H\,{\sc ii} regions and there is no known source of nonthermal
filaments in its vicinity.  However, there is a bright compact radio source
detected at 90~cm with a flux density of 112 mJy at $l$=40$'\ 2.1''$,
$b$=--2$' 7''$ detected at 90~cm (Nord et al.  2004).  This source is
likely to be nonthermal due to colliding winds from binary systems
similar to the bright nonthermal source in the Arches cluster
(Yusef-Zadeh et al.  2007, in preparation).  If we assume that this
central nonthermal radio source has a size about $10''\times 10''$
with $\nu^{-0.7}$ spectrum, the equipartition field is 0.36\,mG with
an electron energy density of $2500\u eV \ut cm -3 $ is estimated.
This also yields an estimated flux of $3.3\ee -6 \u ph \ut cm -2 \ut s
-1 $ assuming $q = 150\u ph \ut erg -1 $.  In light of its large column
density and including a factor of 4 leakage from the nonthermal source
because it is entirely enclosed within the molecular cloud (so the
relevant surface area is $4\pi r^2 $ rather than $\pi r^2$), the
predicted flux is still 50 times lower than observed.  Another
difficulty is that strength of the equivalent width of K$\alpha$ line
emission is roughly two times higher than other Galactic center clouds
with the exception  of the Arches cluster.

Thus, the lack of strong localized nonthermal emission (unlike other
galactic center molecular clouds) makes it difficult to apply the LECR
picture to Sgr B2.  However, the strong three-way correlation between
the distribution of molecular clouds, the K$\alpha$ line emission and
TeV emission from the Galactic center clouds suggest that the same
process should be able to explain these observations.  
 Adopting a
size of 96 arcmin$^2$, source depth 10$'$, and a $\nu^{-0.7}$
spectrum, the required electron energy density, equipartition field
and resulting 90\,cm synchrotron flux are $51\u eV \ut cm -3 $,
0.045\,mG and 18\,Jy respectively.   
Observationally, 
there may be
diffuse nonthermal emission surrounding the Sgr B2 cloud. Figure 15a shows 
a low-resolution VLA continuum image of Sgr B2 at 90cm with FWHM=41.6$''\times22.7''$
against contoure of EW  K$\alpha$ line emission. The integrated flux at 90cm is 
about 17 Jy which is  likely  to be mainly due to thermal emission from HII regions.  
It is difficult to disintangle nonthermal from thermal emission from this region. 
There is also  degree-scale 
diffuse 
feature surrounding Sgr B2 based  on Green Bank observations at 90cm (Law et al. 2007, in preparation). 
However, it is not clear how much of this large-scale diffuse emisison,  which is more than sufficient 
to account for the origin of 6.4 keV emission, arises from Sgr B2. The total background subtracted integrated 90cm 
flux from the region covering Sgr A, B, C and the continuum arc is about $\sim$1200 Jy, most of which 
is expected to be nonthermal.

The large EW of
Fe K$\alpha$ line emission in Sgr B2 may result from the high
abundance of iron by roughly a factor of 3-4 times solar value.  Other
parameters that could increase the strength of EW are column density
and the spectrum of the nonthermal emission, but neither of which can
be important as shown in Figures 13a,b.  The Galactic center molecular 
clouds show metallicity that is generally twice higher than in the 
solar neighborhood, thus we are predicting that Sgr B2 could have a
1.5-2  times higher metallicity than that of typical 
Galactic center clouds. 

Sgr B1 lies adjacent to a number of
nonthermal radio filaments with flux densities of $\sim$10 mJy on the
eastern side of the Sgr B1 cloud, Figure 15b shows the contour
distribution of EW of K$\alpha$ emission against the grayscale 90~cm
image.  The prominent nonthermal filaments observed at 90~cm clearly
shows their distribution with respect to the edge of K$\alpha$ line
emission.  The integrated 90~cm flux from the filaments shown in Figure
15b is roughly about 800 mJy.  The distribution of EW over the region
of Sgr B1 shown in Figure 2 ranges between few hundreds to 1500 eV
near $l$=34$' 7''$ , $b$=--1$' 13''$.  The extracted spectrum over this region
gives a EW$\sim570$\,eV (Table 2).   Adopting a
size of 77 arcmin$^2$, source depth 10$'$, and a $\nu^{-0.7}$
spectrum, the required electron energy density, equipartition field
and resulting 90\,cm synchrotron flux are $23\u eV \ut cm -3 $,
0.030\,mG and 3.2\,Jy respectively.

\subsection{Cosmic Ray Heating of Molecular Clouds}

The heating associated with the electron energy
losses is significant.  It is dominated by the electrons entering
the cloud with initial energies between 0.1 and 1\,MeV as lower energy
electrons are injected less efficiently, and higher energy electrons
do not lose much of their energy in the cloud.  For each erg of
electron energy deposited into the cloud, approximately 10\% goes
into heat (Dalgarno, Yan \& Liu 1999) and 200 Fe K$\alpha$ photons
are produced, thus the heating rate is
\begin{equation}
    \frac{\Gamma}{n_\mathrm{H}} = 0.1 \times \frac{4\pi
    I_{\mathrm{K}_\alpha}}{N_\mathrm{H}} \,
\frac{2/3}{200\,\mathrm{ph\,erg^{-1}}} \sim
    3\ee -24 \u erg \ut s -1 \ut H -1
    \label{eq:Gamma}
\end{equation}
where the factor of $2/3$ removes the contribution of $E>10\u MeV $
electrons to $I_{K_\alpha}$.  This heating rate implies temperatures
of $\sim 200\u K $ for $n(\mathrm{H}_2)\sim 1\ee 4 \ut cm 3 $ and
$N_\mathrm{H_2}/\Delta v \sim 10^{22} \ut cm -2 \ut km -1 \u s $
(Neufeld, Lepp \& Melnick 1995).  Such a high temperature due to low
energy cosmic ray heating is consistent with multi-transition ammonia
line observations indicating a warm 200K gas observed throughout the
Galactic center region (e.g., H\"uttemeister et al.  1993b;
Rodriguez-Fernandez et al.  2004; Oka et al.  2005).  There is also a
denser and colder component of molecular gas that has been detected to
have small volume filling factor.  If the denser component lies in the
Galactic center, we believe that either cosmic rays have not
penetrated these dense regions or that the higher cooling rate of
dense gas has lowered the gas temperature.

\subsection{Cosmic Ray Ionization Rate in the Galactic center}

We can also estimate that the ionization of the molecular gas
associated with the losses of electrons with incident energies
between 0.1 and 1\,MeV. The ionization rate is readily estimated from
eq (11) by noting that on average, an ionization occurs for each
deposition of 40.1\,eV of electron energy into the gas (Dalgarno
\etal\ 1999), yielding $\zeta \approx (3\ee -24
\u erg \ut s -1 \ut H -1 ) / (40.1\u eV ) \approx 5\ee -13 \ut s -1 \ut H -1
$.  This value of $\zeta$ is about
4 orders of magnitude higher than the canonical value of the
interstellar cosmic-ray ionization rate and corresponds to an
energetic electron energy density $\sim 1000\u eV \ut cm -3 $,
appropriate for Sgr C. This is reduced by 1--2 orders of magnitude
for Sgr B1 and B2 and the Radio Arc (see Table 3), giving $\zeta \sim
2 \ee -14 \ut s -1 \ut H -1 $ for these sources.  As pointed out
earlier, the detection of substantial amount of the (3,3) metastable
rotational levels of H3$^+$ in the Galactic center region requires an
estimated high ionization rate which is estimated to be about two
orders of magnitude higher than in that for diffuse clouds (Oka et al.
2005).  Thus cosmic-rays in the Galactic center region may be
responsible for producing the large reservoir of warm molecular gas
which is inferred from H3$^+$ measurements.

\section{The Origin of Diffuse TeV Emission}

Inverse Compton scattering of the sub-millimeter radiation from dust in the
Galactic centre clouds by relativistic electrons may account for or
significantly contribute to the diffuse TeV emission observed towards
the central regions of the Galaxy (Aharonian et al.\ 2006).  For
example the strongest HESS peak overlays the arc (see Fig. 12).  
To estimate this contribution,  
we assume that the
dust emission in the region is isotropic with intensity $I_\nu =
B_\nu(T_d)(1-\exp(-\tau_d))$ where $T_d = 20\,K$ is the dust temperature
and the dust optical depth $\tau_d\propto\nu^2$ (Pierce-Price et al.\
2000).  The 850$\micron$ and 400$\micron$ SCUBA maps yields an
estimate of $\tau_d=0.029$ at $\nu = kT_d/h$.  The electron spectrum
associated with the arc must extend to at least 30 TeV to produce
upscattered photons at TeV energies.  An extrapolation of the $E^{-2}$
electron spectrum that obtains at GeV energies into the TeV range
predicts a photon flux at 1 TeV that is sixty times that observed, which
we estimate to be $\sim 1\ee -13 \u ph \ut cm -2 \ut s -1 $ (Aharonian
et al.\ 2006).  However one expects a synchrotron break at lower
energies, and the observed TeV spectrum has a photon index of $\sim 
-2.3$, steeper than the $-1.5$ this simple model would predict.  Instead we
assume that synchrotron losses steepen the electron spectrum at
energies above 2\,TeV to an $E^{-3}$ dependence; the corresponding
synchrotron loss time scale is $1\,000$ years.  This steepens the
predicted ICS photon spectrum from $E_\gamma^{-1.5}$ to
$E_{\gamma}^{-2}$ above 40\,GeV and matches the observed HESS flux at
1\,TeV.

\section{Conclusions}

The distributions of \thco\, line, X-ray, nonthermal radio continuum and 6.4 keV line emission 
suggest against the irradiation origin of X-ray emission from Sgr C. We believe that the origin 
of X-ray continuum and line emission can naturally be explained by the interaction of cosmic rays 
with prominent Galactic center molecular clouds such as Sgr C, the ``45 and -30 \kms clouds'', 
Sgr B1 and Sgr B2. We have argued that the sources of nonthermal particles are due to a 
combination of the Galactic center radio filaments which are localized in the immediate vicinity 
of molecular clouds, as well as the diffuse nonthermal component distributed throughout the 
Galactic center region. This hypothesis is consistent with similar value of 6.4 keV equivalent 
width throughout the Sgr~C, the -30 \& 45 \kms clouds near the radio arc and Sgr B1 regions.

In the context of this model, we explain the high value of the equivalent width of Fe K$\alpha$ 
line and unusually strong K$\alpha$ line emission from Sgr B2 in terms of higher metallicity and 
the contribution of a diffuse nonthermal particles in this source. The interaction of the 
nonthermal radio filaments and the molecular cloud implies that these features are at the same 
distance from each other and that the long-standing problem of the origin of high temperature 
molecular gas can naturally be explained in terms of the LECR heating of the clouds. In addition, 
we estimate the high cosmic-ray ionization rate in the Galactic center region
which could explain 
the origin of warm molecular gas as traced by H$_3^+$ absorption study. 
In addition, 
we provide an alternative model to explain the origin of diffuse TeV emission from molecular 
clouds in the central region of the Galaxy by using the high energy component of nonthermal 
electrons to upscatter the submillimeter emission from dust clouds in this region. The remarkable 
correlation of K$\alpha$ line emission, TeV emission and molecular gas suggest that similar 
processes are at work in explaining both diffuse X-ray and $\gamma$-ray emission from the 
Galactic center clouds.
Lastly, we believe  that the interplay between the cosmic rays and  neutral gas in the 
central  region of the 
Galaxy is  significant and provides much insight in our understanding of the physical 
processes that are involved in both the  nuclei  and the disk of galaxies.


\acknowledgements
Acknowledgments: We are grateful to J. Hinton, N. Kassim, D. Pierce-Price \& Y. Tsuboi for 
providing us  with their published data. D. Lis is supported by NSF grant
AST-0540882 to Caltech Submillimeter Observatory. 

\section{Appendix}

\subsubsection{Polarization  Measurements}

Additional support for the synchrotron nature of the filament comes
from polarization measurements.  We carried out high frequency
polarization study of Sgr C at 3.6~cm to reduce the effect of Faraday
rotation toward this source.  Figure 10a shows a grayscale image of
linearly polarized emission from the linear filament of Sgr C with
line segments indicating the polarization angle distribution at 3.6~cm.
The peak emission has a flux density of 1.4 mJy beam$^{-1}$ near the
center of the image and $\sim$0.28 mJy beam$^{-1}$ away from the
center.  The degree of polarization is remarkably high corresponding
to a value of about 65\%.  In contrast, the region to the south of
this highly polarized feature, the polarization degree drops off to
10\% near the compact source G359.45--0.05.  The low degree of linear
polarization to the south of the filament could be due to the
contamination by the bright thermal emission from Sgr C H\,{\sc ii} region.
This diffuse H\,{\sc ii} region is likely responsible for depolarization of
synchrotron emission from the Sgr C filaments.  Due to high Faraday
rotation toward the Galactic center sources, we are not able to
determine the intrinsic direction of the magnetic field along the Sgr
C filament.  Assuming that depolarization is caused by an external
medium, the lack of strong polarized emission to the south suggests
that the filament lies either behind the H\,{\sc ii} region or is embedded
within it.

Figure 10b shows a grayscale image of Sgr C at 3.6~cm with a resolution 
9.8$''\times7.5''$ (PA=--7.5$^{\circ})$. 
We note  several circular-like 
extended features to the north and south of the Sgr C filament, as well as a 
bright compact source along the filament. These features  generally 
 show a flatter spectrum than 
the  linear filament. The compact source G359.45--0.05 
($l$=$359^{\circ} 27' 19.94''\pm0.002$, $b$=$-03' 22.49''\pm0.04$) 
has   a spectral index of  
$\sim -0.29\pm 0.21$ between 3.6 and 2~cm. Other extended sources identified 
as 
G359.45--0.07, G359.45--0.03 and G359.45--0.02 in Figure 9, all of which appear to be 
thermal sources having 
mean spectral index values $\alpha=-0.1\pm0.24, -0.09\pm0.3$ and --0.02$\pm0.3$, 
respectively.  It is difficult to determine the true relationship between these 
features and the nonthermal filament with the available resolution. 
In particular,  
the ring-shaped morphology of G359.45--0.07, in which the nonthermal filament 
appears to pass through the ring, as well as 
the compact source G359.45--0.5. 
We also note a weak vertical filament to the south of the 
Sgr C H\,{\sc ii} region directed south of galactic latitude $b$=--6$'$, as shown in Figure 10b. 
This feature 
needs to be confirmed as it  falls outside the 
primary beam of our 3.6~cm image and that it has never been detected  in previous  
radio continuum images of Sgr C.


\clearpage

\begin{figure} 
\epsscale{.8}
\plotone{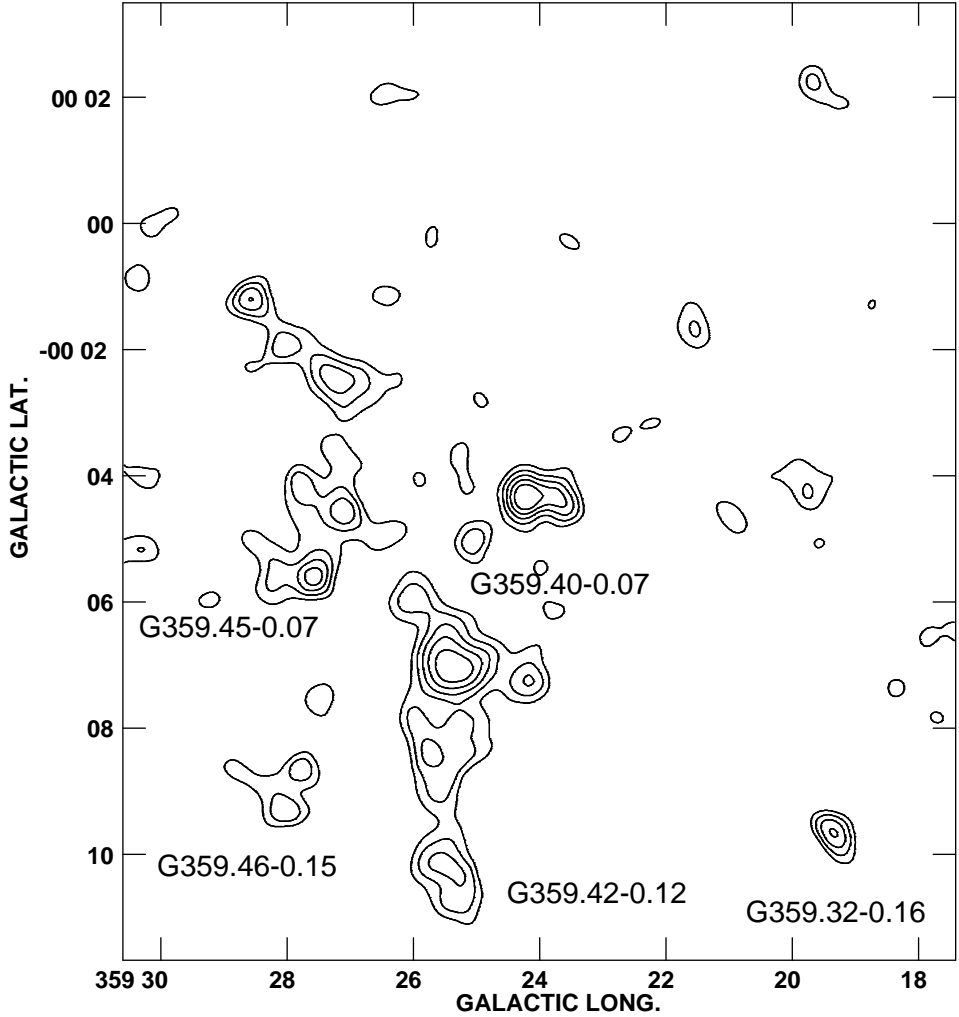} 
\caption{{\bf (a)} Contours of X-ray
continuum emission between 2 and 6 keV. The X-ray data are convolved with a
Gaussian beam having FWHM$\sim30''$. Contour levels are set at (7, 8, 9, 10,
11)$\times10^{-9}$ Jy per beam. The diffuse and compact X-ray continuum
features are labeled.  {\bf (b)} The top and bottom panels show modeling of the
X-ray spectrum of diffuse emission from Sgr~C and the corresponding residuals
in the bottom panel.  The thick solid line is the model, the dashed and the
thin solid lines correspond to the 5 and 1 keV plasma components, the
dot-dashed line is the emission line from low-ionization iron, and the
dot-dot-dot-dashed line is the power law. The residuals are in units of the
1-sigma uncertainty on the spectrum.  As shown in Figure 2, the spectrum was
taken from an ellipse centered on $l$=359.43382$^{\circ}$,
$b$=--0.09655$^{\circ}$ with a radii of 4.1\arcmin\ and 2.9\arcmin\ along the
$l$ and $b$ directions, respectively (PA=28.6$^{\circ}$).
}
\end{figure}  
\clearpage
{\plotone{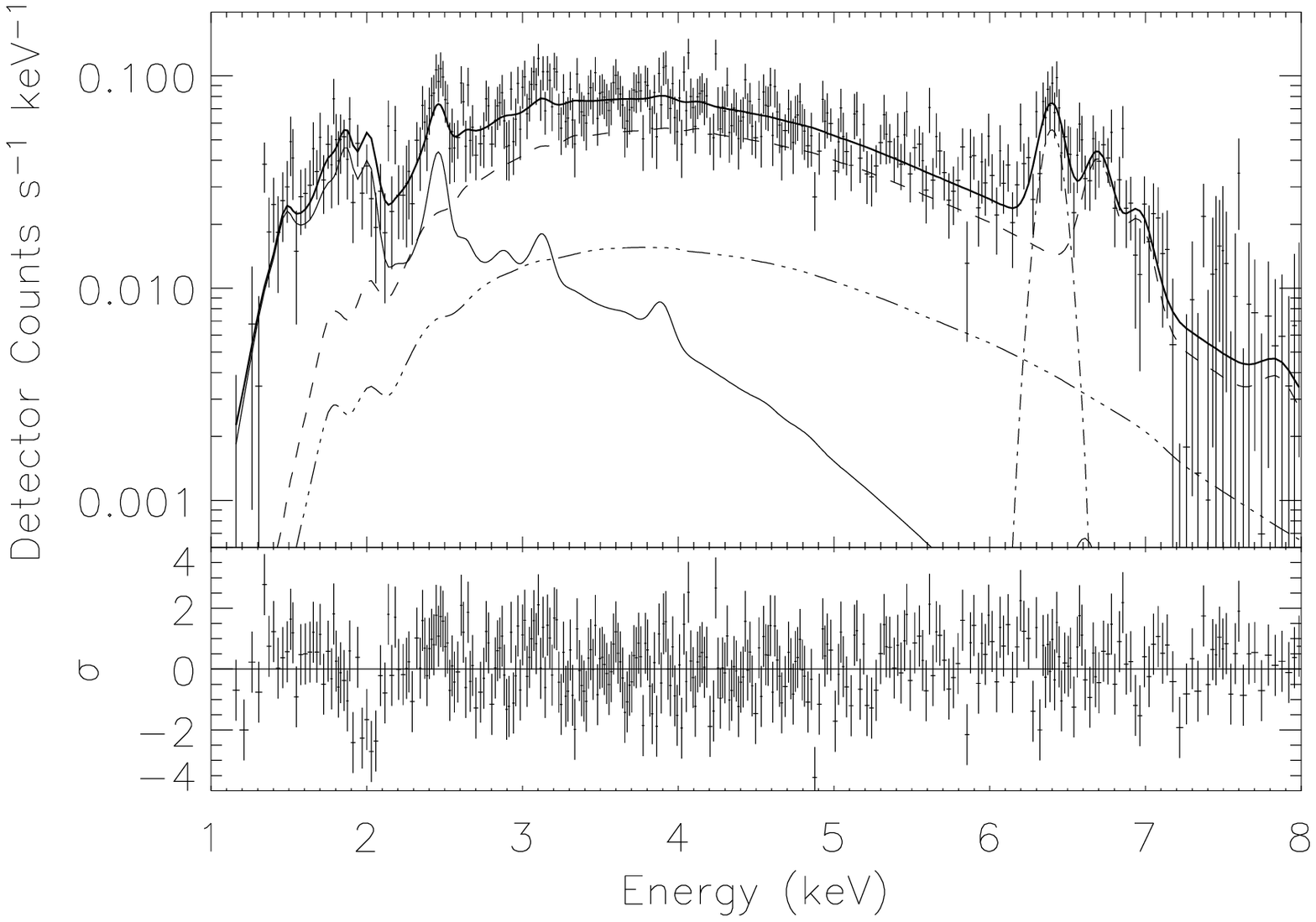}}\\[5mm]
\centerline{Fig. 1b. ---}
\clearpage
\begin{figure}
\includegraphics[scale=0.6,angle=0]{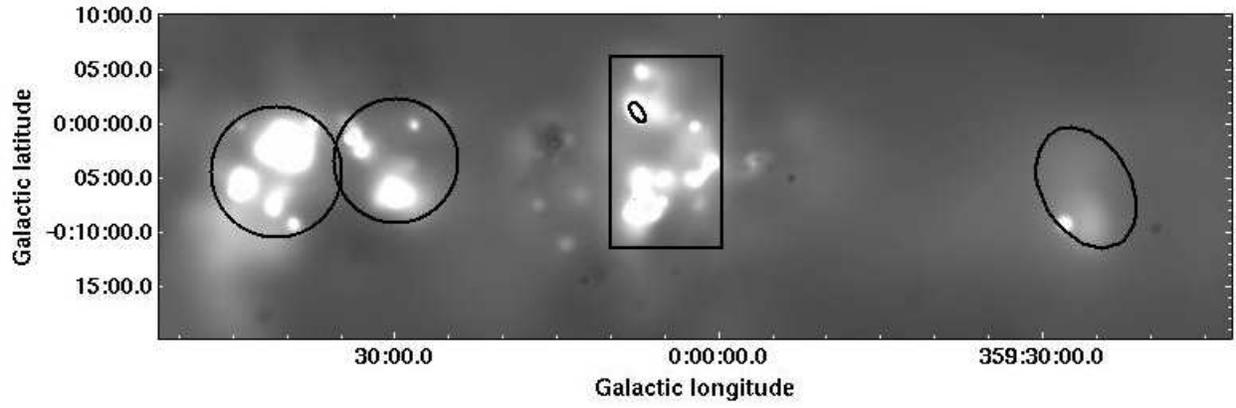}
\caption{{\bf } 
A  distribution of  K$\alpha$  6.4 keV EW line emission. The regions from 
which X-ray spectra are  extracted are drawn as an ellipse (Sgr C, the Arches cluster), 
a rectangle (the arc) and circles (Sgr B1 and Sgr B2).}
\end{figure}  

\begin{figure}
\plotone{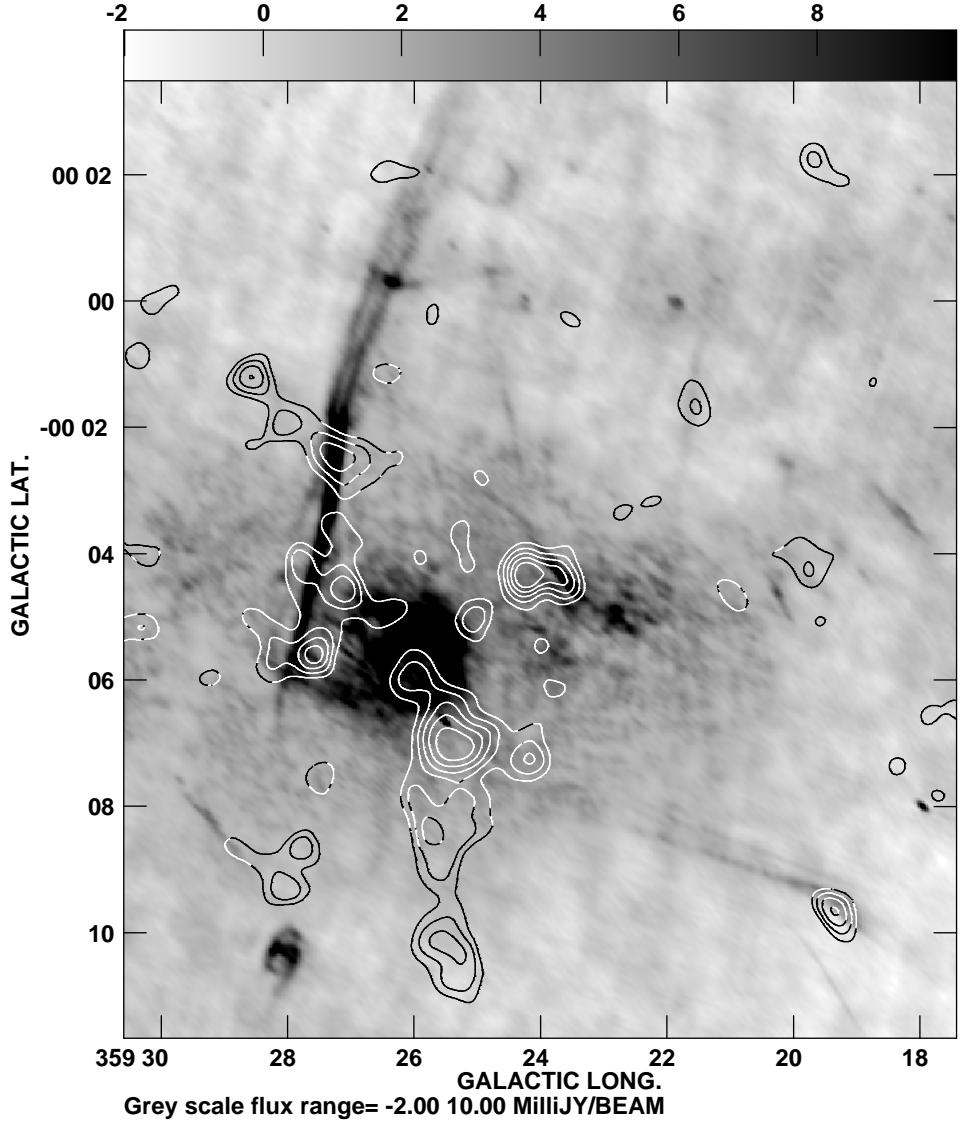}
\caption{
{\bf (a)}
 Contours of X-ray emission between 2 and 6 keV are 
superimposed on  a 20~cm continuum image with a resolution of 
9.8$''\times4.2''$ (PA=--12$^{\circ}$).  The contour  levels are set 
at (7, 8, 9, 10, 11) $\times10^{-9}$Jy.
The X-ray data are convolved by a  Gaussian having FWHM=30$''^2$. 
{\bf (b)} Contours of X-ray emission are the same as in (a). 
The 
surface brightness  of a 20~cm  continuum image  with a 
resolution of 30$''^2$ ranges   
between -9 and 200 mJy beam$^{-1}$ 
in order to
bring out the  faint features. 
{\bf (c)} Contours of X-ray emission are the same as in (a). 
The 
surface brightness  of a 90~cm  continuum image  with a 
resolution of 12.6$''\times6.8''$ (PA=3$^{\circ}$) taken from 
Nord et al. (2004). 
{\bf (d)} 
A close-up view of G359.32--0.1  with  contours of 
X-ray emission  superimposed 
on a 20~cm continuum image (in grayscale) with a spatial resolution of 
8.1$''\times3.3''$ (PA=--11$^{\circ}$). 
The contour  levels are set 
at 0.8, 1, 1.2, 1.4  nJy.
The X-ray image is convolved to a  9$''$ Gaussian. 
}\end{figure}  
\clearpage
{\plotone{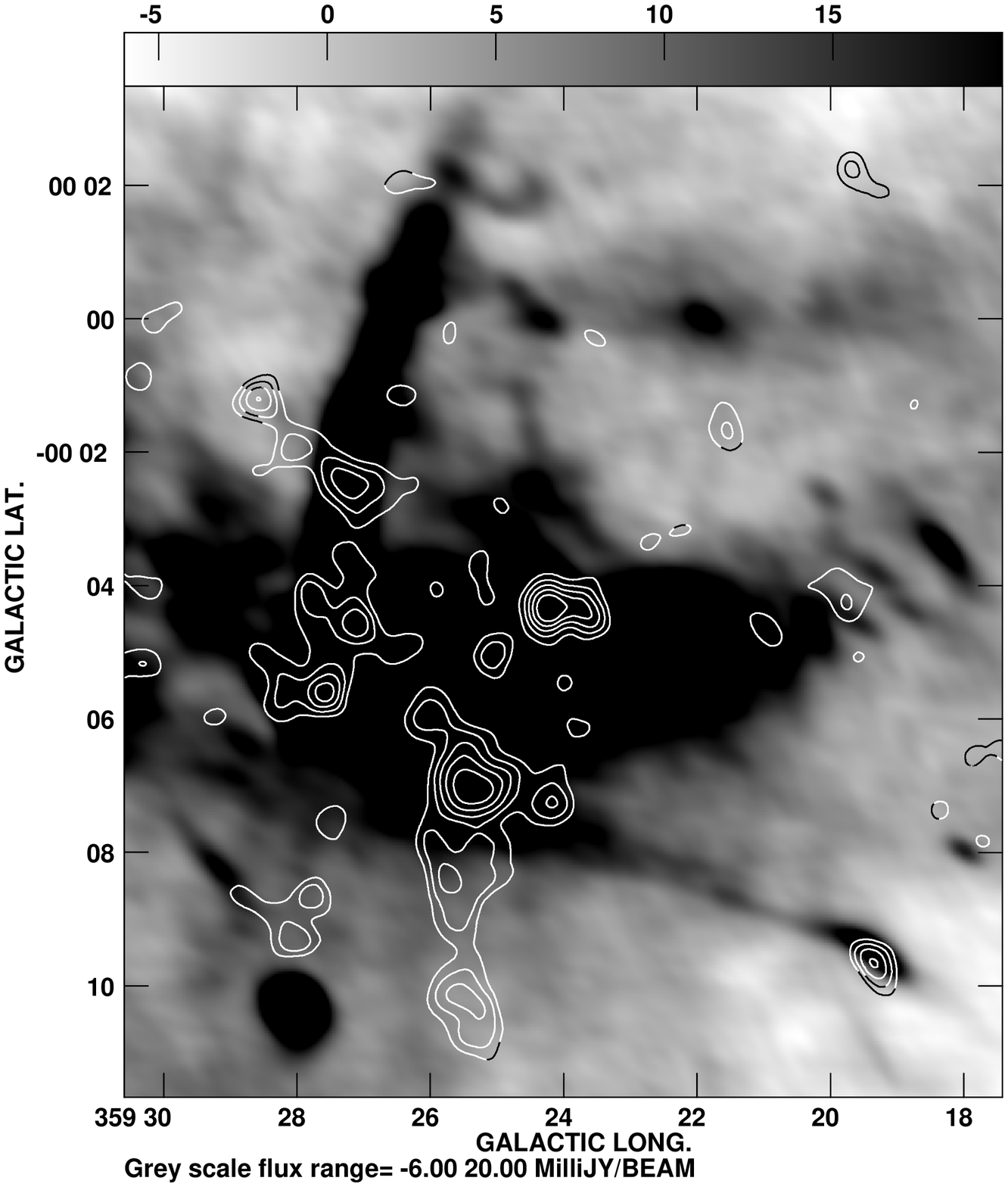}}\\[5mm]
\centerline{Fig. 3b. ---}
\clearpage
{\plotone{f3c.ps}}\\[5mm]
\centerline{Fig. 3c. ---}
\clearpage
{\plotone{f3d.ps}}\\[5mm]
\centerline{Fig. 3d. ---}
\clearpage
\begin{figure}
\begin{center}
\includegraphics[width=0.95\textwidth, clip]{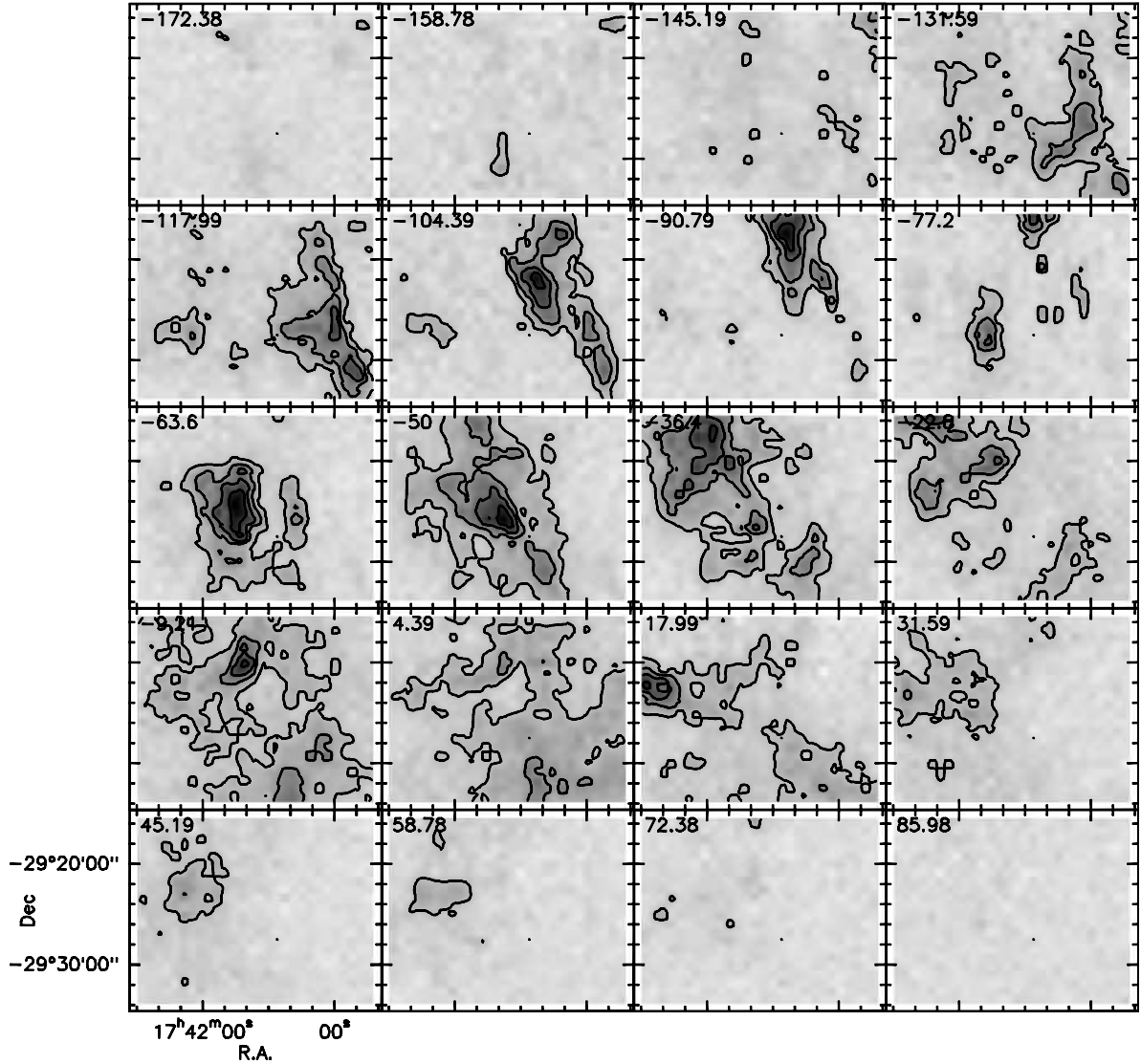}
  \caption{\thco~(1--0) velocity channel maps. Channel center
  velocities are given in the upper-left center of each panel. The
  emission associated with the Sgr~C cloud is seen in the central part
  of the map in the --77 to --36~\kms\ panels. Contour levels are
  --0.5, 0.5, 1, 1.5, and 2~K. The emission has been corrected for
  the atmospheric attenuation and warm losses, but not for the main
  beam efficiency ($\eta = 0.4$). This figure 
is shown in J2000 celestial coordinates. 
  \label{fig:chan}}
\end{center}
\end{figure}

\begin{figure}
\epsscale{.9}
\plotone{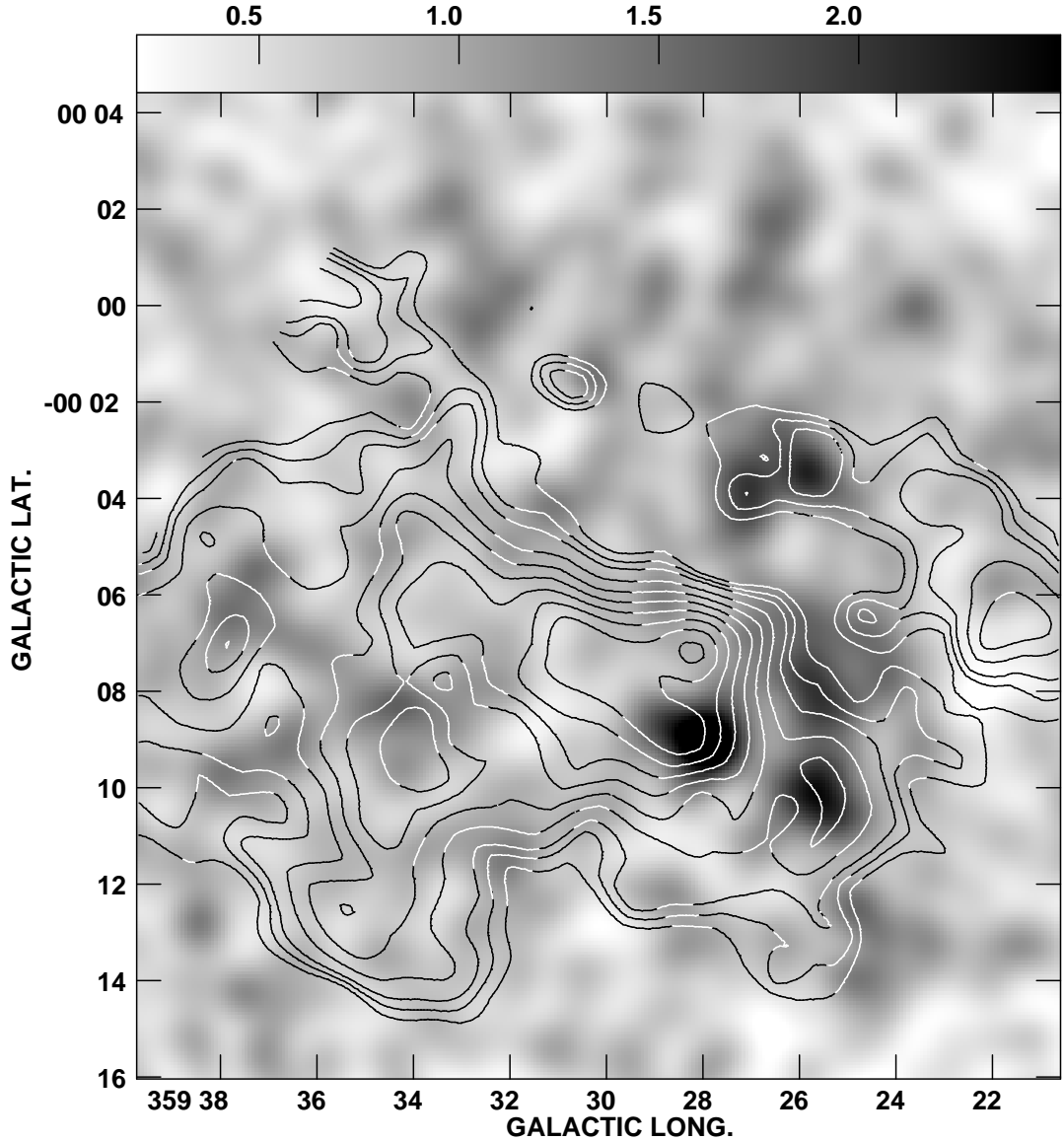}
\caption{{\bf(a)} 
Contours of velocity
integrated $^{13}$CO emission  between
-77.2 and -22  \kms are superimposed on a grayscale 6.4 keV  
K$\alpha$ line emission in grayscale 
with a 30$''$ resolution. 
 Contour levels are set at
(2, 2.25, 3, 3.5, 4, 4.5, 5, 5.5, 6) $\times$ 14 K\, \kms{}.
{\bf (b)} 
 Similar to (a) except that 
the grayscale (grayscale) shows the continuum X-ray emission 
with 60$''$ 
spatial resolution between 
2 and 6 keV.
{\bf (c)} 
Contours of X-ray continuum emission
are superimposed on a grayscale 
equivalent width map of 6.4 keV line emission.}
\end{figure}  
\clearpage
{\plotone{f5b.ps}}\\[5mm]
\centerline{Fig. 5b. ---}
\clearpage
{\plotone{f5c.ps}}\\[5mm]
\centerline{Fig. 5c. ---}
\clearpage

\begin{figure}
\plottwo{f6a.ps}{f6b.ps}
\caption{{\bf (a)} A plot  of the spectral index 
distribution along  the 
brightest nonthermal vertical filament of Sgr C.  
 {\bf (b)} Similar to (a) except that the slice plot shows  
1$\sigma$ error bars of the spectral index distribution. 
These measurements are  based 
on the 20 and 90~cm images convolved to a  Gaussian beam  having 
FWHM=12.6$''^2$. }
\end{figure}  
\clearpage
\begin{figure}
\epsscale{1}
\plotone{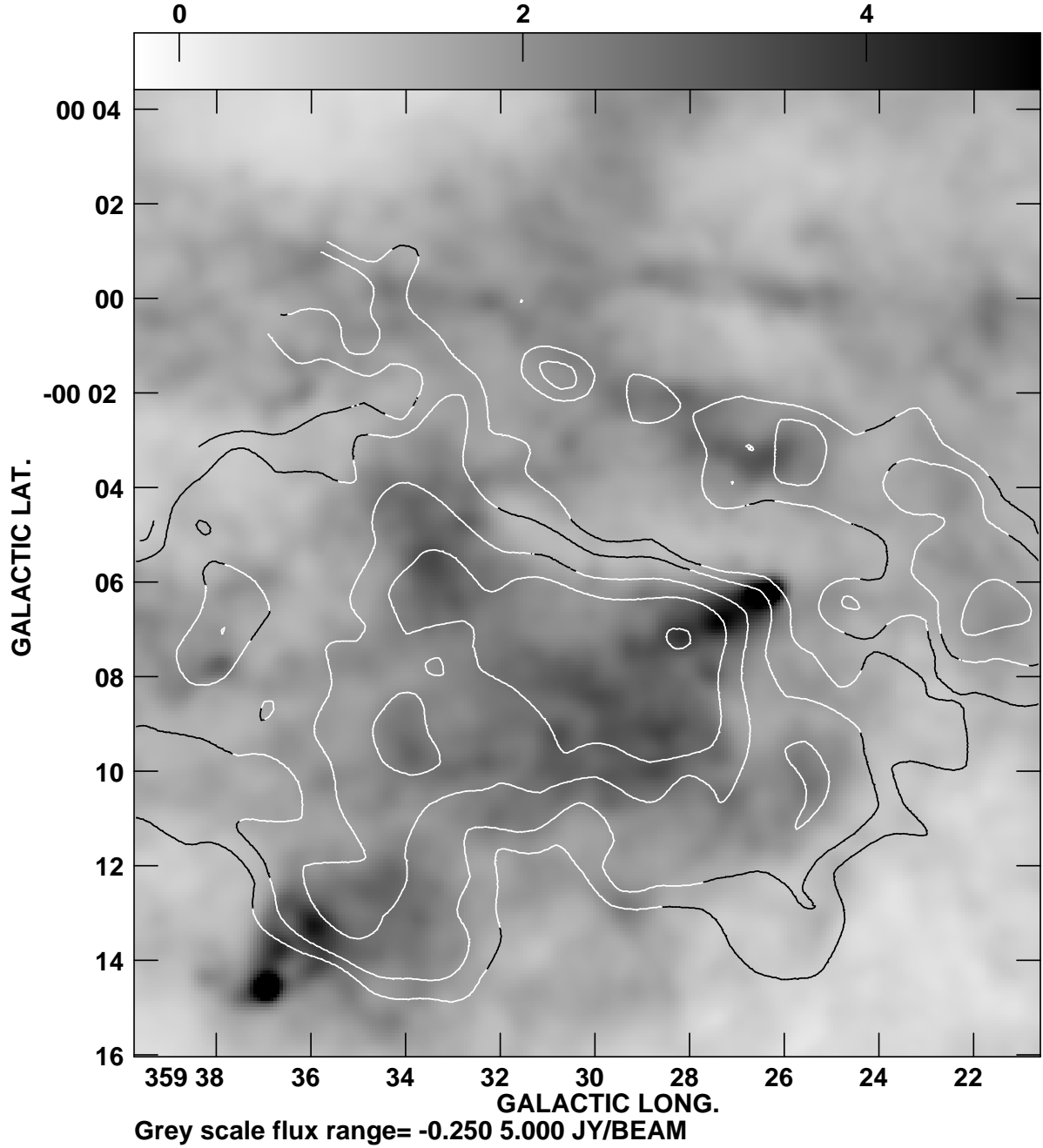}
\caption{{\bf } Contours of \thco\ molecular line emission are
superimposed on a grayscale sub-millimeter image at 850$\mu$m taken from 
Pierce-Price et al. (2000) convolved to a resolution of 15$''$. }
\end{figure}  

\begin{figure}
\plottwo{f8a.ps}{f8b.ps}
\caption{{\bf (a)} A plot  of the spectral index 
distribution along  the 
brightest nonthermal vertical filament of Sgr C.  
 {\bf (b)} Similar to (a) except that the slice plot shows  
1$\sigma$ error bars of the spectral index distribution. 
These measurements are  based 
on the 20 and 90~cm images convolved to a  Gaussian beam  having 
FWHM=12.6$''^2$. }
\end{figure}  

\begin{figure}
\plottwo{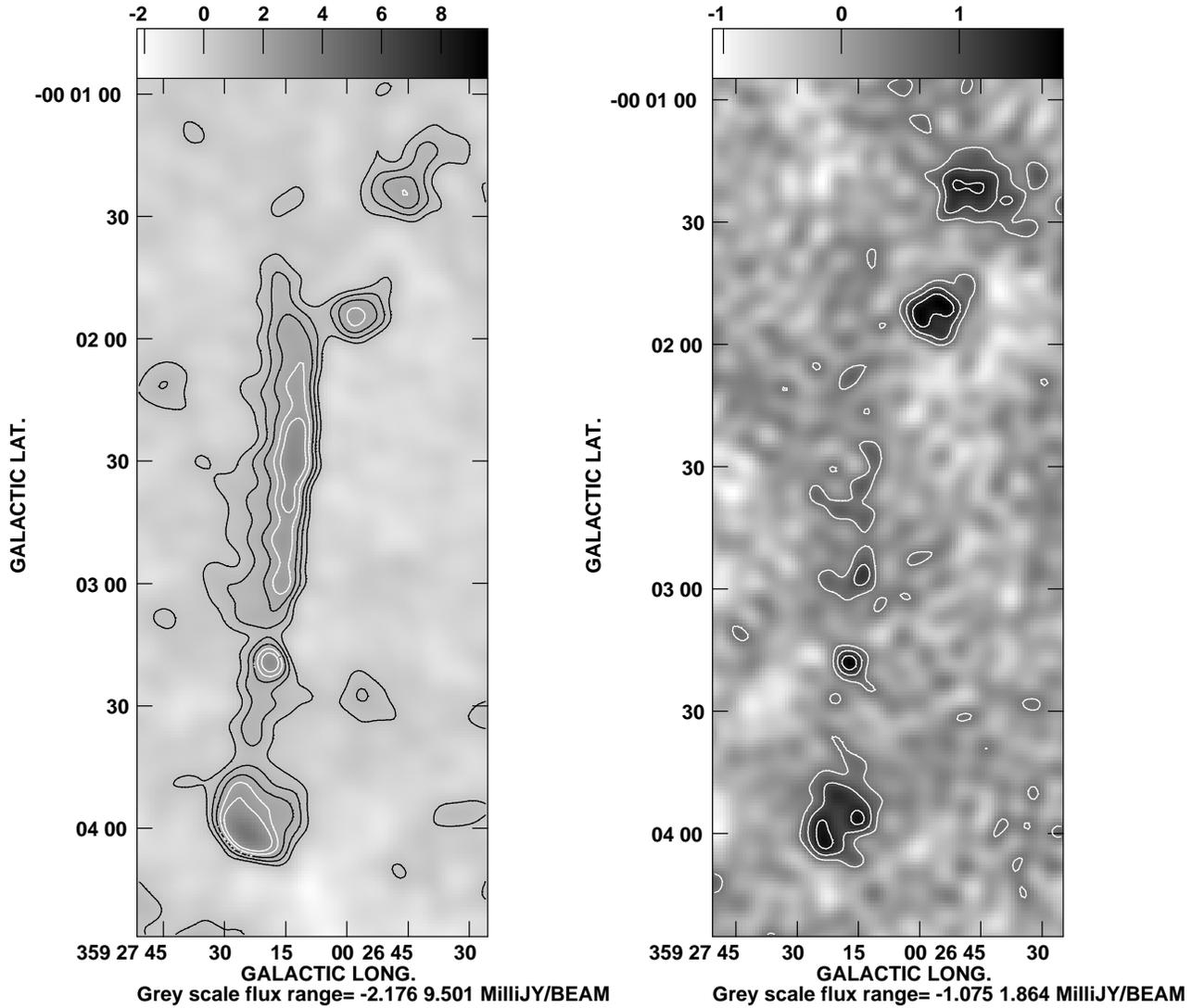}{f9b.ps}
\caption{{\bf (a) Left} A grayscale  continuum 
image of the bright filament of 
Sgr C at 3.6~cm with contours set at 0.25 $\times$ (2, 4, 6, 8, 10) Jy
 beam$^{-1}$. 
The rms  noise is 0.24 mJy and the spatial resolution is $6.8''\times6.4''$ 
(PA=10.2$^{\circ}$).  
 {\bf  (b) Right}  Identical to (a) except that 2~cm contours are set at 0.25 
$\times$ 
(2, 4, 6) Jy beam$^{-1}$.  
Both images are  constructed by using the same {\it {uv}} range (2 -- 30 
k$\lambda$) to  insure the same spatial frequency coverage.}
\end{figure}  

\begin{figure}
\includegraphics[scale=0.6,angle=0]{f10a.ps}
\includegraphics[scale=0.6,angle=0]{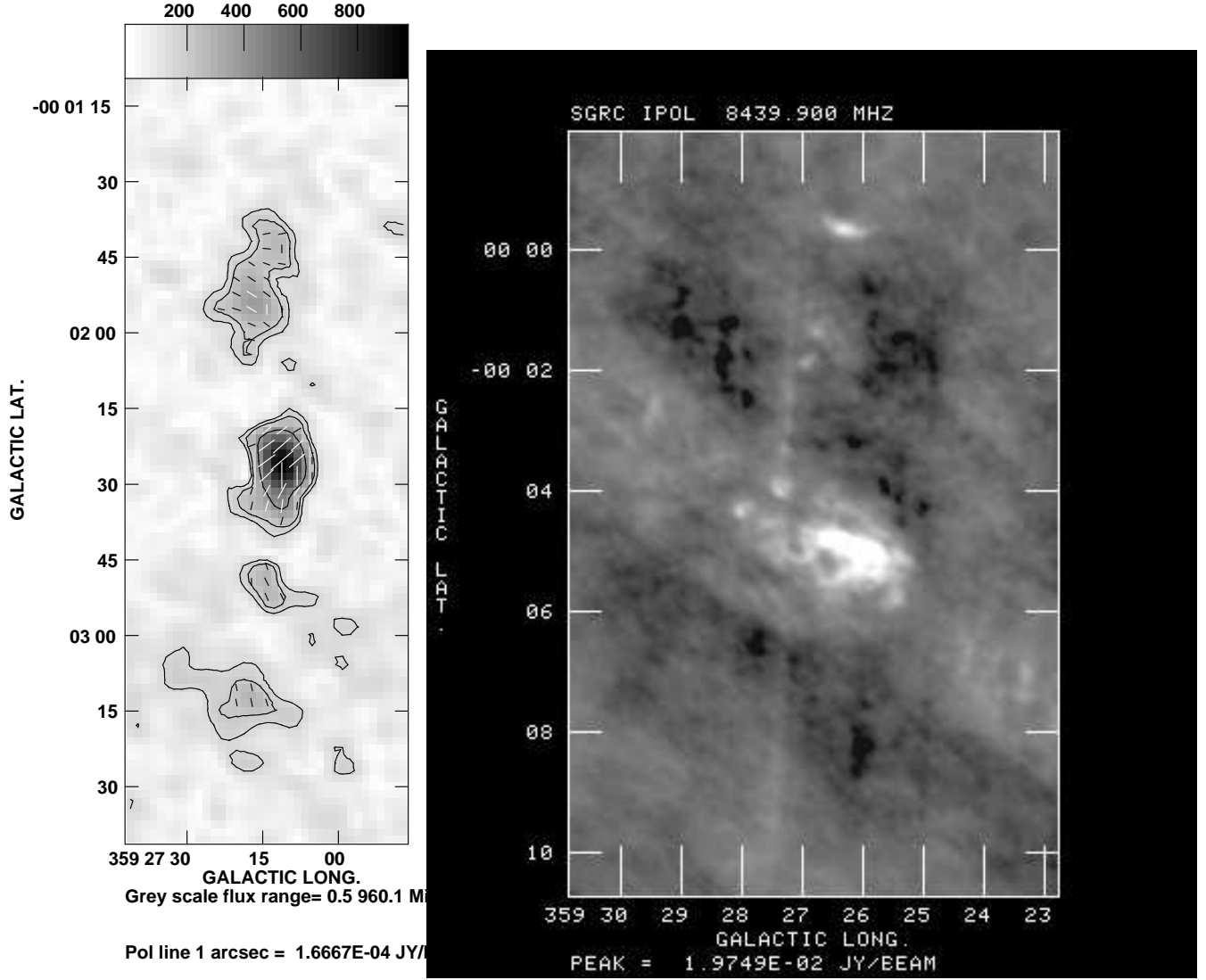}
\caption{{\bf (a) Left}
A grayscale linearly polarized continuum emission 
from the bright filament of Sgr C at 3.6~cm 
 with contours set at 0.15, 0.2, 0.4 mJy 
beam$^{-1}$ and the resolution of $7.9''\times6.7''$ (PA=--6.6$^{\circ}$). 
The length of the line segments  presents the strength
 of the  polarized emission 
(1$''$ corresponds to  0.166 mJy beam$^{-1}$) 
whereas the position angle of the line segments shows the 
polarization angle distribution.
{\bf (b) Right} A grayscale image of Sgr C at 3.6~cm with a resolution of 
$9.8''\times7.5''$ (PA=--7.5$^{\circ}$). This image is not corrected for the 
response of the primary beam. The weak filament to the south lies outside the 
FWHM of the primary beam. }
\end{figure}

\begin{figure}
\plotone{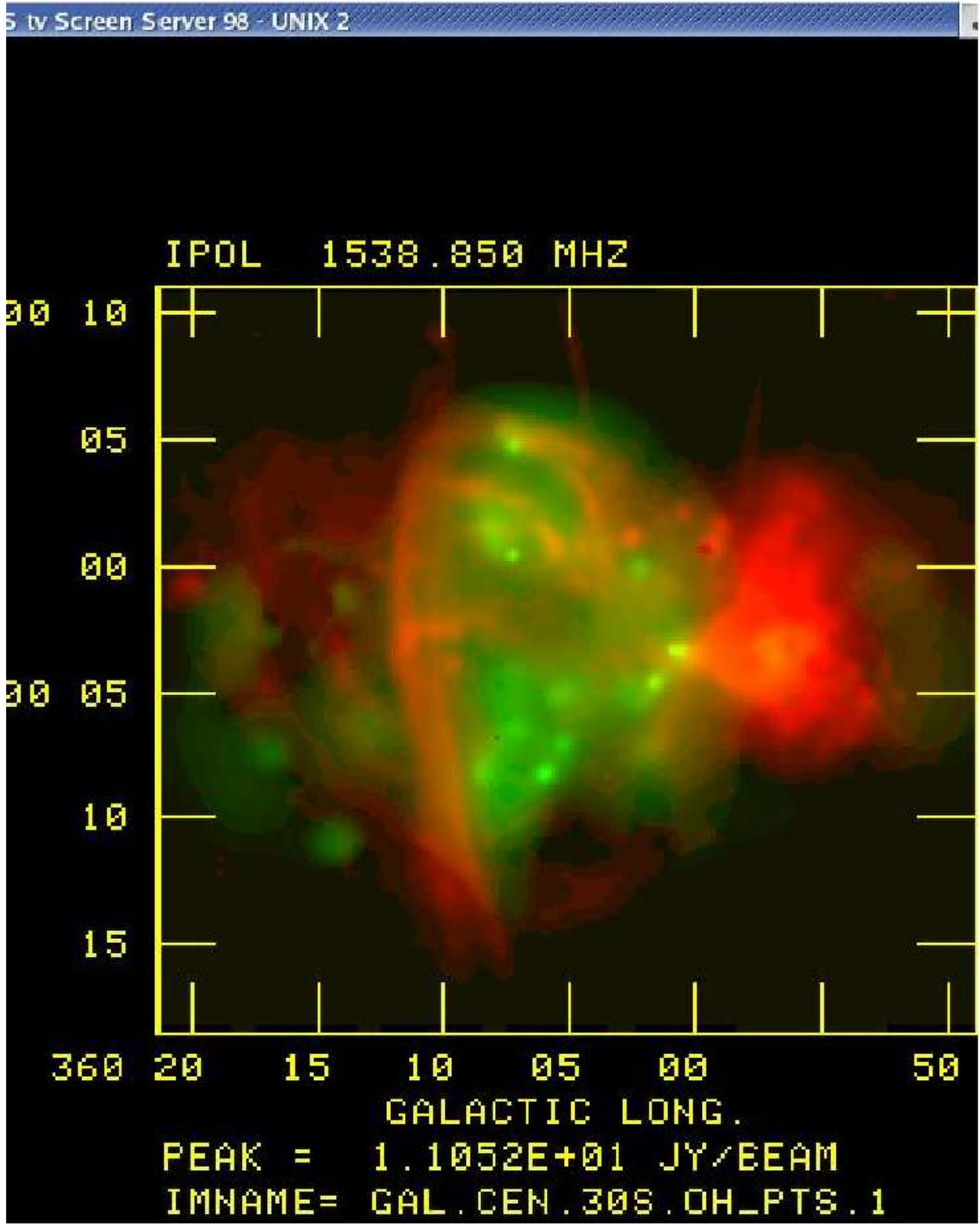}
\caption{{\bf (a)} A 20~cm continuum  image of the
Sgr A complex and the radio arc in Galactic  coordinates  (red
color) is
shown against the distribution of the 6.4 keV K$\alpha$ line emission
(green  color). {\bf (b) } A 20~cm continuum  image of the
Sgr A complex and the radio arc in Galactic  and celestial coordinates 
against contours  of ammonia line emission from prominent molecular clouds in 
the Galactic center region (G\"usten et al. 1981; Yusef-Zadeh 1986).
{\bf (c)} A grayscale image of the 6.4 keV line EW distribution with 
contours set at 1, 1.5, 2, 2.5, 3, 3.5, 4, 5, 6, 8, 10, 14, 18)$\times10^2$ eV. 
The crosses show the ammonia peaks , a shown in (b) and the five-pointed stars show the 
peak position of C line emission from Serabyn and G\"usten 1987).    }
\end{figure}
\clearpage
\centerline{\includegraphics[scale=1,angle=0]{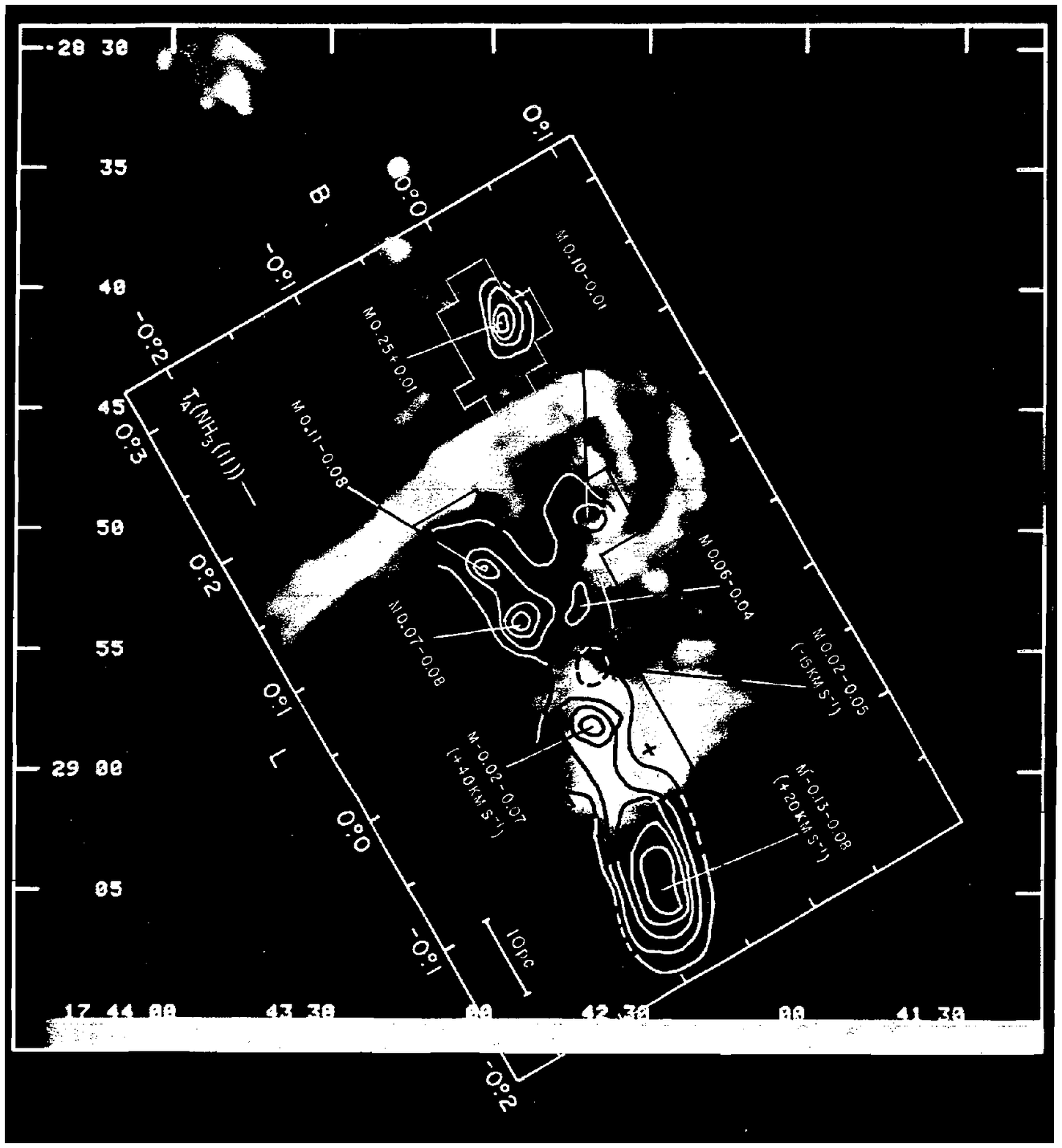}}
\centerline{Fig. 10b. ---}
\clearpage
\centerline{\includegraphics[scale=0.8,angle=-90]{f11c.ps}}
\centerline{Fig. 11c. ---}
\clearpage
\begin{figure}
\includegraphics[scale=0.7,angle=-90]{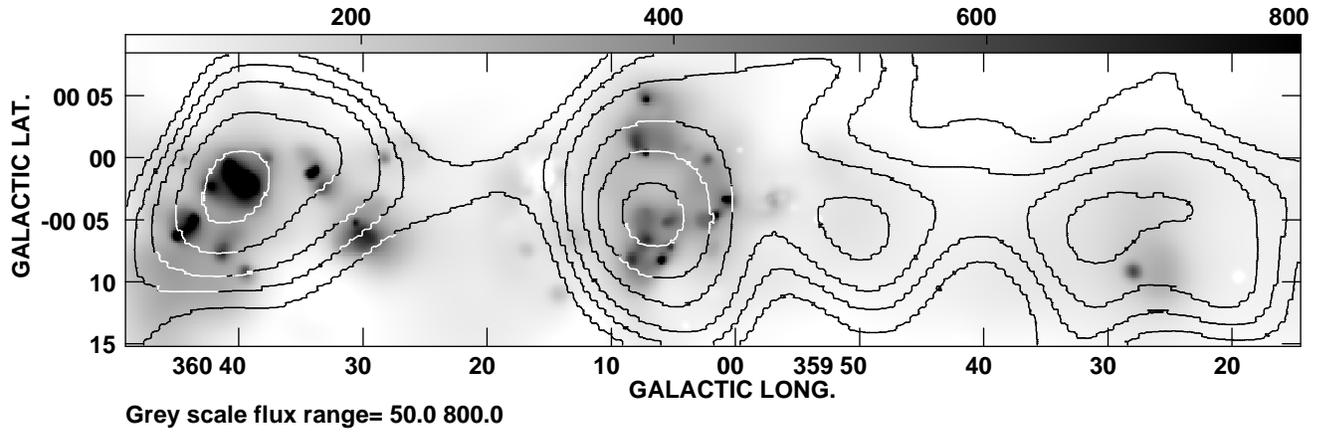}
\includegraphics[scale=0.9,angle=0]{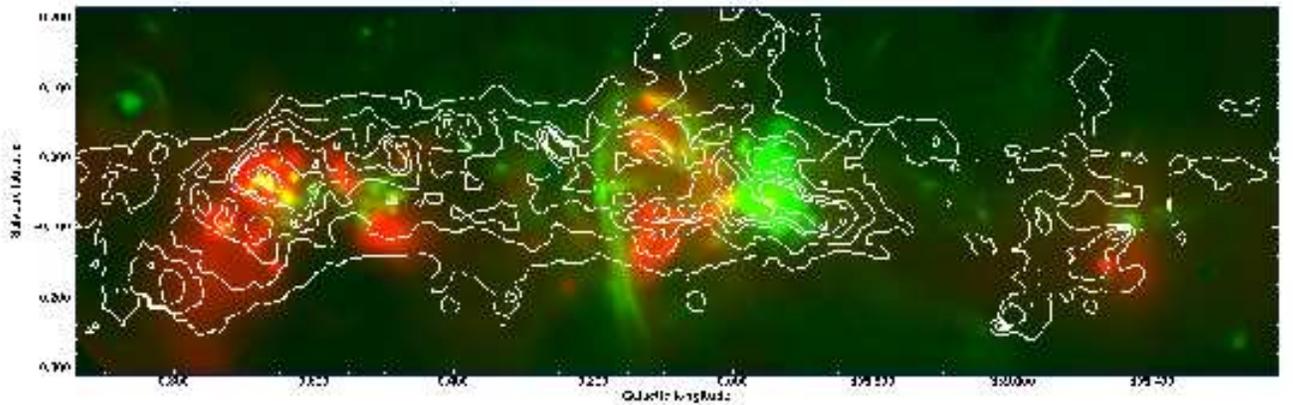}
\caption{{\bf (a)} Contours of HESS emission  from the Galactic center (Aharonian et al. 
2006) is superimposed on the distribution of  K$\alpha$  6.4 keV EW line emission. 
{\bf (b)} Contours of 850 micron  submillimeter emission are superimposed
on the distributions of  K$\alpha$  6.4 keV EW line emission (in red) 
and a 20~cm continuum emission (in green color).}
\end{figure}  

\vfill\eject
\begin{figure}
\includegraphics[scale=1,angle=0]{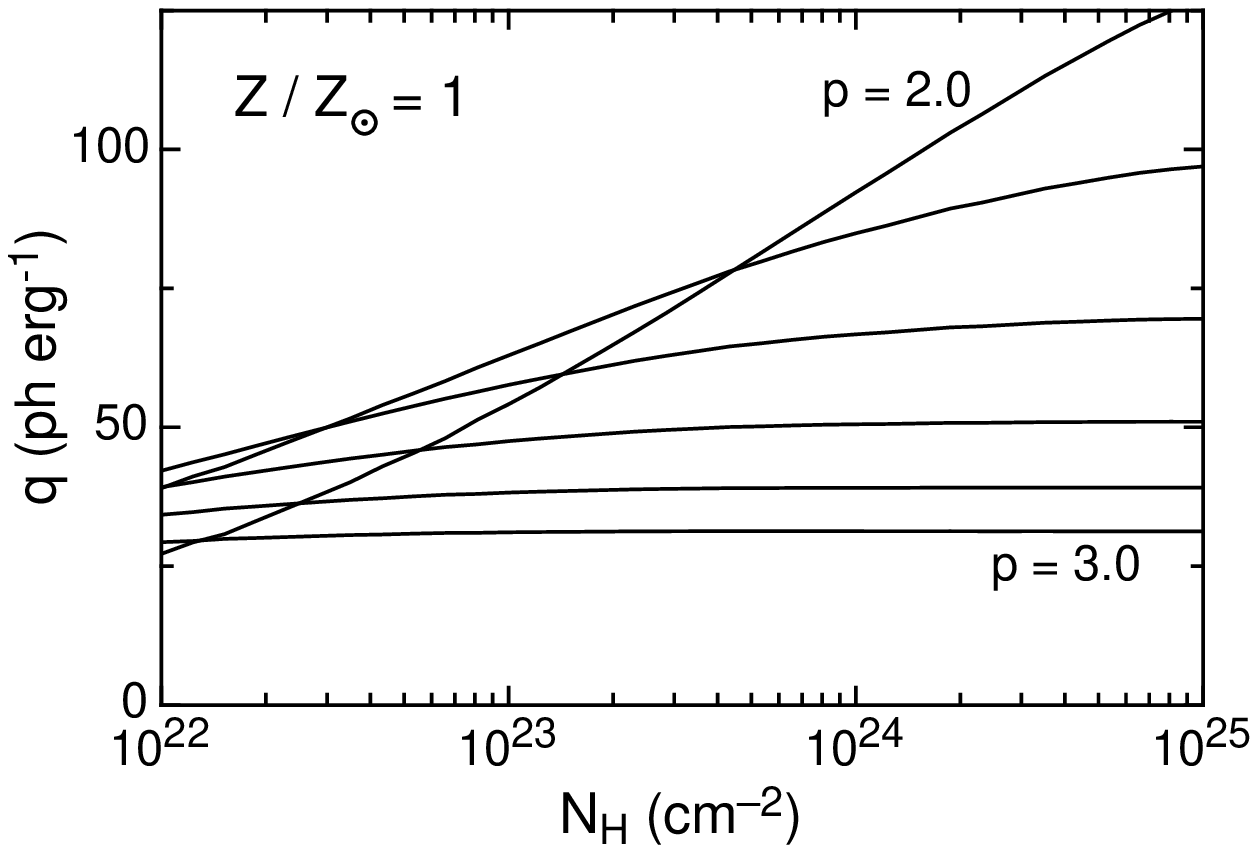}
\includegraphics[scale=1,angle=0]{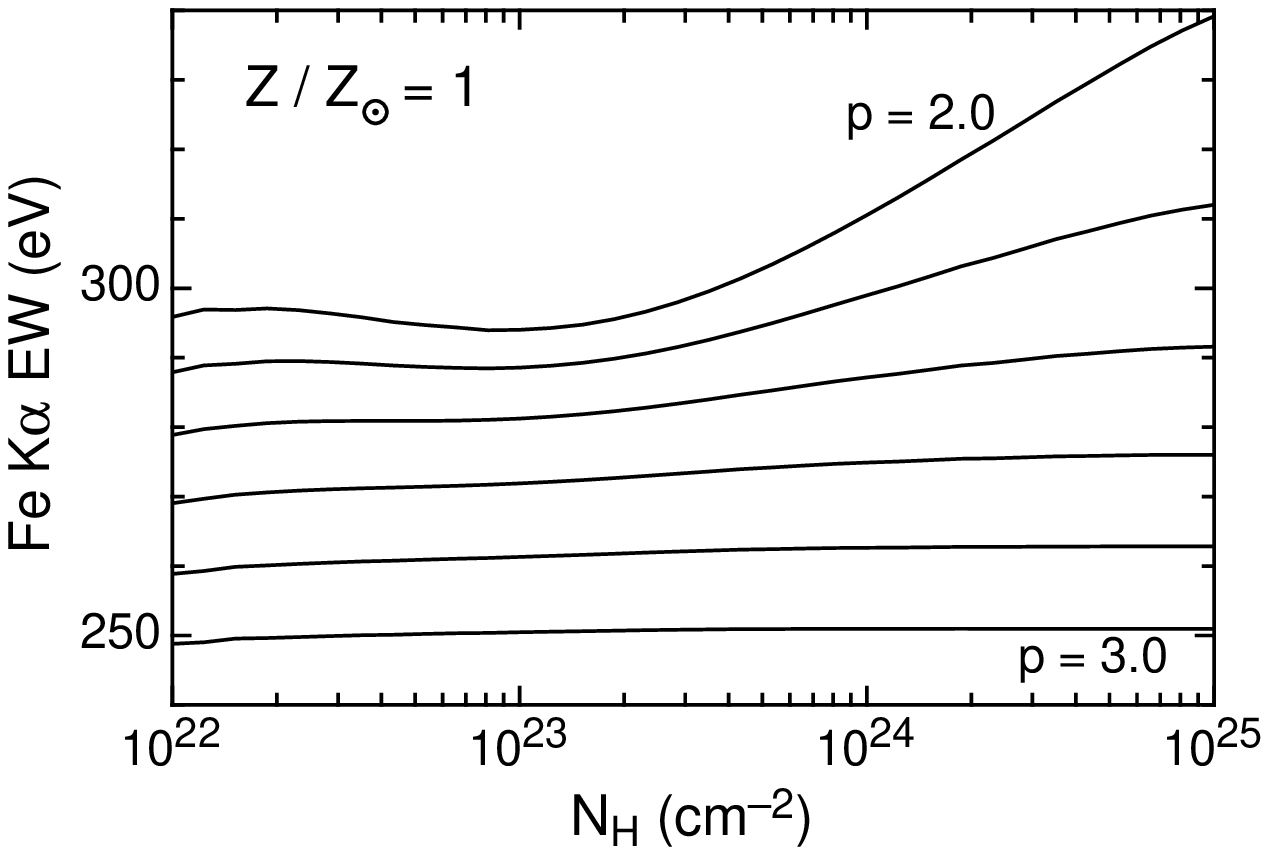}
\caption{{\bf a} Fe K$\alpha$ production per erg of electron energy injected 
into a cloud of given column density $N_\mathrm{H}$ and a solar iron 
abundance (Fe/H = $2.8\ee -5 $).  The curves are 
labeled by the power-law index $p$ of the electron energy spectrum 
($\propto E^{-p}$), which is assumed to run from 10\,keV to 1\,GeV.
{\bf b}  Similar to (a) except that the 
equivalent width of Fe K$\alpha$ is shown as a function 
of 
column density for different values of the particle spectral index.}
\end{figure}

\vfill\eject
\begin{figure}
\includegraphics[scale=0.75,angle=0]{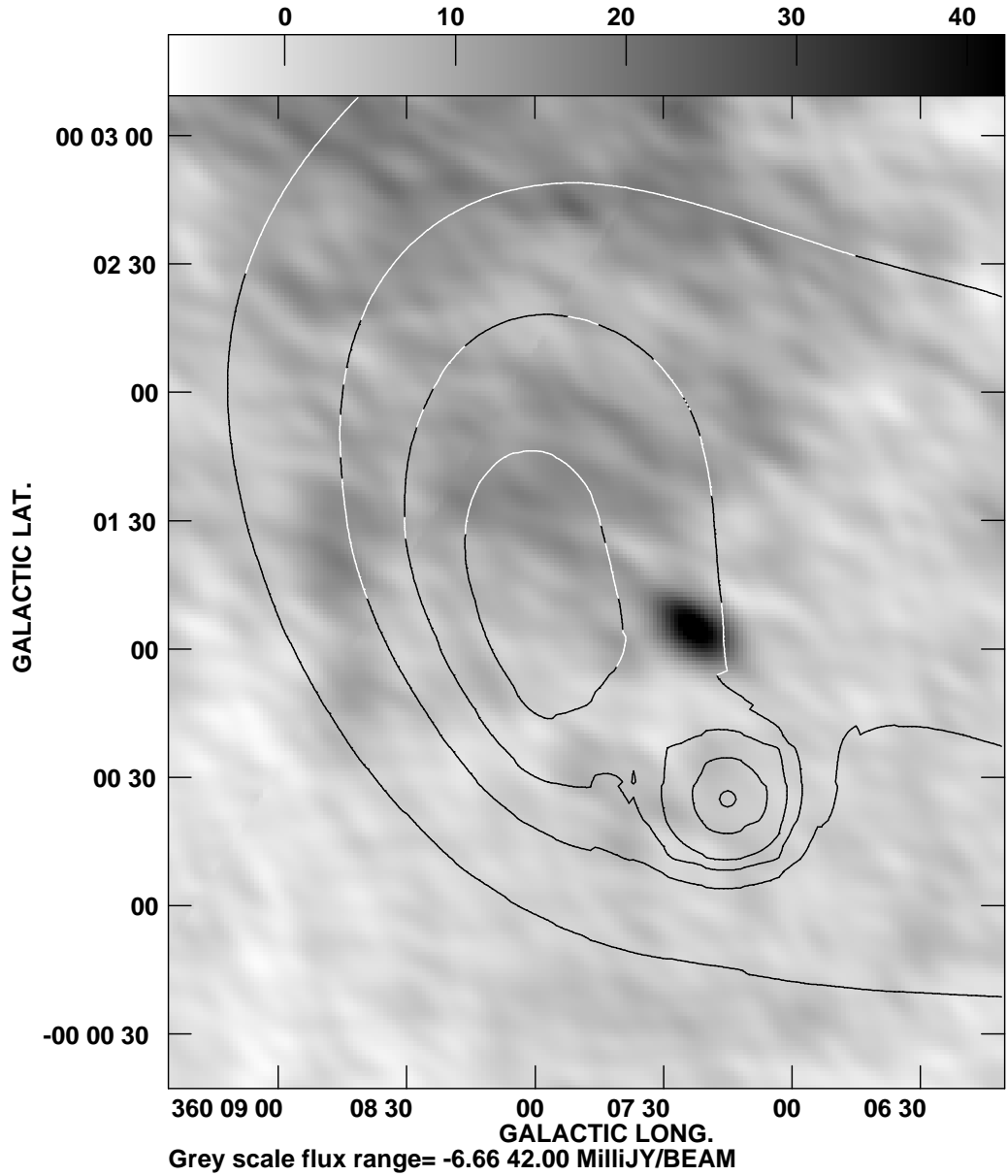}
\caption{Contours of EW of Fe K$\alpha$ line emission  with levels
(3, 4, 5, 6, 8, 10) $\times$ 100 eV are 
superimposed on a grayscale distribution of 
90~cm emission from the Arches cluster (in black color).}
\end{figure}

\begin{figure}
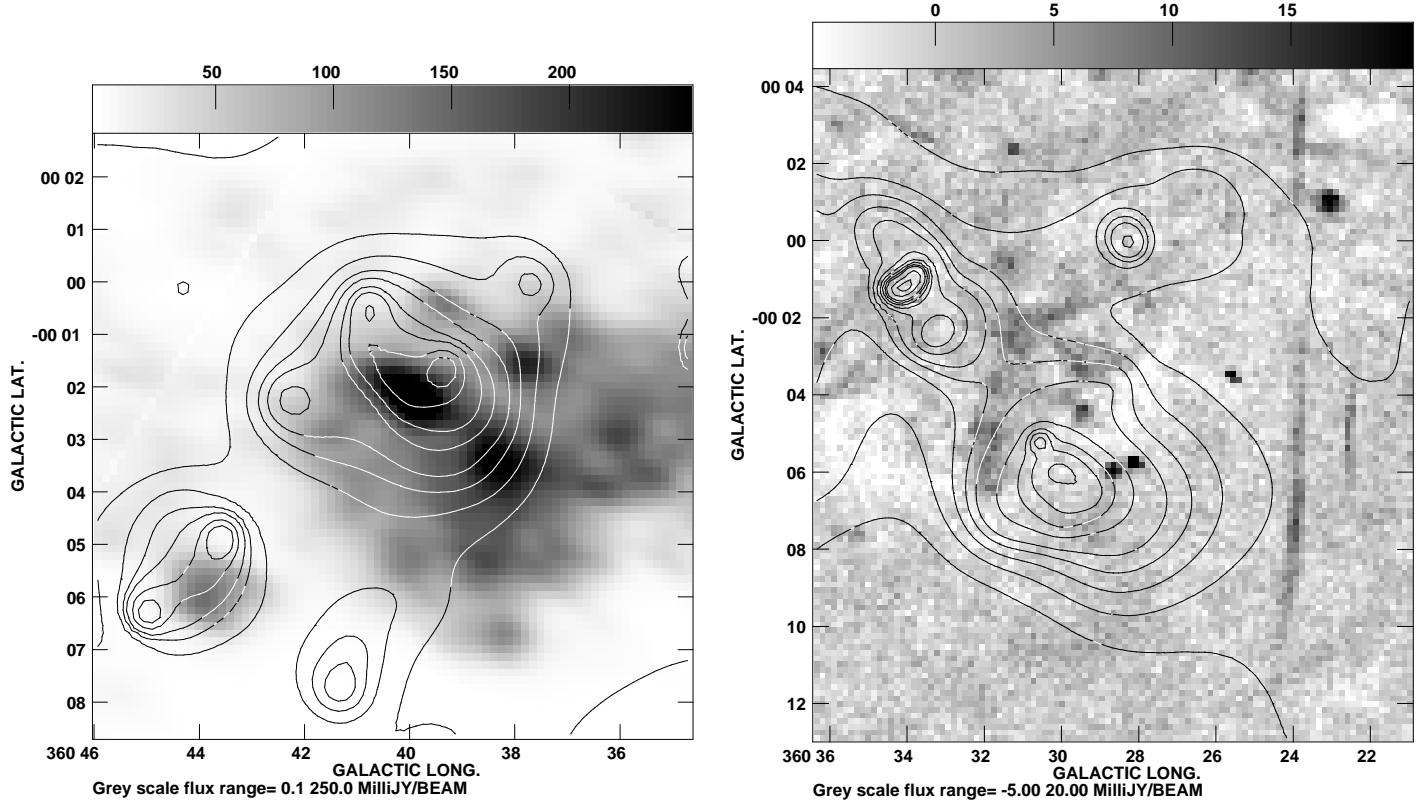

\includegraphics[scale=0.5,angle=0]{f15a.ps}
\includegraphics[scale=0.5,angle=0]{f15b.ps}
\caption{{\bf a} Contours of EW of Fe K$\alpha$ line emission from Sgr B2 with levels
(1, 2,  3, 4, 5, 6,  8, 10, 14, 18, 22) $\times$ 100 eV are 
superimposed on a grayscale distribution of 90~cm emission from Sgr B2 with a resolution of 
$41.7''\times22.7''$  (PA=-6$^0$) (Anantharamaiah et al. 1991).
{\bf b} 
Contours of EW of Fe K$\alpha$ line emission from Sgr B1 with levels
(1, 1.5, 2, 2.5, 3, 4, 5, 6, 7, 8, 11, 14) $\times$ 100 eV are 
superimposed on a grayscale distribution of 90~cm emission from Sgr B1 and G0.6--0.0 
in black color with a resolution of 12.6$\times6.8''$ (PA=$3^0$).
}
\end{figure}


\begin{thebibliography}{99}


\bibitem[]{}
Anantharamaiah, K. R.,  Pedlar, A.,  Ekers, R. D. \& Goss, W. M.
1991, MNRAS, 249, 262	

\bibitem[]{1872}
Aharonian, F. et al. 2006, Nature, 439, 695
\bibitem[Arnaud \etal(1996)]{arn96} Arnaud, K.A., 1996, Astronomical Data 
 Analysis Software and Systems V,  eds. Jacoby G. and Barnes J., p17, ASP 
 Conf. Series volume 101

\bibitem[]{1878}
Anantharamaiah, K.R. \& Yusef-Zadeh, F. 1989, in the center of the Galaxy, IAU Symp. 136.  ed.
M. Morris (Kluwer Academic Publisher, Dordrecht), p159

\bibitem[]{1882}
Armstrong, J.T. \& Barrett, A.H. 1985, ApJ, 57, 535


\bibitem[]{1886}
Belanger, G., Goldwurm, A., Renaud, M., Terrier, R., Melia, F.,
Lund, N., Paul, J., Skinner, G. \& Yusef-Zadeh, F. 2006, ApJ, 636, 275

\bibitem[]{}
Boldyrev, S. \& Yusef-Zadeh, F. 2006, ApJ, 637, L101

\bibitem[]{1892} Caswell, J.L. \& Haynes, R.F.  1987, A\&A, 171, 261

\bibitem[]{1894}
 Coil, A. \& Ho, P.T.P. 2000, ApJ, 533, 245
                                                                                        

\bibitem[]{1898}
Cotera, A., Erickson, E., Colgan, S., Simpson, J., Allen, D.
\& Burton, M. 1996, ApJ, 461, 750
                                                                                     
                                                                                     

\bibitem[]{1904} Dalgarno, A., Yan, M. \& Liu, W. H. 1999,
         ApJSupp, 125, 237

\bibitem[] Downes, D.,  Goss, W. M., Schwarz, U. J. \&  Wouterloot, J. 
G. A. 1979, A\&AS, 35, 1	

\bibitem[] Ekers, R. D.,  van Gorkom, J. H.,  Schwarz, U. J. \&  Goss, 
W. M. 1983, A\&A, 122, 143	


\bibitem[Erickson et al. (1992)]{erickson92} Erickson, N.R., Goldsmith,
  P.F., Novak, G., Grosslein, R.M., Viscuso, P.J., Erickson, R.B., \&
  Predmore, C.R. 1992, IEEE Trans, Microwave Theory Tech. 40, 1


\bibitem[]{1912}
Figer, D.F., Sungsoo, S.K., Morris, M., Serabyn, E.,
Rich, R.M., McLean, I.S. 1999, ApJ, 525, 750
                                                                                     

\bibitem[]{1917}
Frerking, M. A.,  Langer, W. D. \&  Wilson, R. W. 1982,  ApJ,  262, 59

\bibitem[]{1920}
Fryer, C. L., Rockefeller, G., Hungerford. A. \& Melia, F. 2006, ApJ, 638, 786

\bibitem[]{1923}
Giveon, U.,  Morisset, C. \& Sternberg, A. 2002, A\&A 392, 501


\bibitem[]{1927}
Gray, A.D. 1994, MNRAS, 270, 861

\bibitem[]{1930}
G\"usten, R., Walmsley, C.M. \& Pauls, T. 1981, A\&A, 103, 197

\bibitem[]{1933}
Handa, T., Sakano, M., Naito, S. \& Hiramatsu, M. 2006, ApJ,  636, 261

\bibitem[] Helfand, D. J. \&  Becker, R. H. 1987, ApJ, 314, 203

\bibitem[]{1936}
Ho, P.T.P., Jackson, J.M., Barrett, A.H. \& Armstrong, J.T., 1985,
ApJ, 288, 575
                                                                                        

\bibitem[]{1941}
H\"uttemeister, S.,  Wilson, T.L., Banina, T.M., Martin-Pintado,
J.  1993b, A\&A, 280, 255

\bibitem[]{1945}
H\"uttemeister, S., Wilson, T., Henkel, C. \& Mauserberger, R. 1993a, A\&A,
 276, 445


\bibitem[]{1950} ICRU 1984,
International Commission on Radiation Units and Measurements Report 37

	
\bibitem[] Kassim, N. E. \& Frail, D. A. 1996, MNRAS, 283, L51

\bibitem[]{k96} Koyama,  K., Maeda, 
Y., Sonobe, T., Takeshima, T.,
Tanaka, Y.   \etal  1996, PASJ, 48, 249

\bibitem[]{1958}
Lang, C.C., Goss, W.M. \& Morris, M. 2002, AJ, 124, 2677

\bibitem[]{1961}
Lang, C.,  Morris, M. \&  Echevarria, L. 1999, ApJ, 526, 727

\bibitem[]{1964} Law, C. 2006, PhD thesis, Northwestern University


\bibitem[]{1967} Law, C. \& Yusef-Zadeh 2004, ApJ, 611, 858



\bibitem[]{1971} Lis, D. C.\& Carlstrom, J. E. 1994, ApJ, 
424, 189	


\bibitem[Lis \& Goldsmith (1989)]{lis89} Lis, D.C., \& Goldsmith, P.F.
  1989, \apj\ 337, 704

\bibitem[Lis \& Goldsmith (1991)]{lis91} Lis, D.C., \& Goldsmith, P.F.
  1991, \apj\ 369, 157


\bibitem[]{1982} 
Liszt, H.S. 1992, ApJS, 82, 495
                                                                                
\bibitem[]{1985} Liszt, H.S., \& Spiker, R.W. 1995, ApJS, 98, 259
                                                                                
\bibitem[]{1987}
Lu, F.G., Wang, Q.D. \& Lang, C.C. 2003,  AJ, 126, 
319       

\bibitem[]{1991} 
Mehringer, D.M.,  Yusef-Zadeh, F.,  Palmer, P.,  \& Goss,
W.M. 1992, ApJ, 401, 168


\bibitem[]{1997} 
Murakami,  H., Koyama, K., Tsujimoto, M. \& Maeda, 
Y.  2001a, ApJ, 550, 297


\bibitem[]{2002} Murakami,  H., Koyama, K. \& Maeda, Y. 2001b, 
ApJ, 558, 687

\bibitem[]{2006} 
Murakami,  H., Koyama, K., Sakano, M., Tsujimoto, M., Maeda, 
Y.  2000, ApJ, 534, 284

\bibitem[Muno \etal(2006)]{mun06} Muno, M. P., Bauer, F.E., Bandyopadhyay, 
R.M. \& Wang, Q.D. 2006, ApJS, 165, 173
 

\bibitem[Muno \etal(2004a)]{mun04a} Muno, M. P. \etal\ 2004a, 
\apj, 613, 326

\bibitem[Muno \etal(2004b)]{mun04b} Muno, M. P. \etal\ 2004b, 
\apj, 613, 1179



\bibitem[]{2018} Neufeld, D. A., Lepp, S. \& Melnick, G. J. 1995,
         ApJ Supp, 100, 132


\bibitem[]{2022}
Nord, M.E., Lazio, T.J.W., Kassim, N.E., Hyman, S.J., LaRosa, T.N., 
Brogan, C.L. \& Duric, N., 2004, AJ, 128, 1646


\bibitem[]{2027}Odenwald, S. F. \& Fazio, G. G. 1984, ApJ, 283, 601

\bibitem[]{2029}
Oka, T., Geballe, T.R., Gotto, M., Usida, T. \& McCall, B. 2005, ApJ, 632, 882

\bibitem[]{2032} 
Oka, T., Hasegawa, T., Sato, F., Tsuboi, M. \& Miyazaki, A.
 2001, PASJ, 53, 779

\bibitem[Park \etal(2004)]{par04} Park, S., Baganoff, F. K., 
Morris, M., Maeda, Y., Muno, M. P., Howard, C., Bautz, M. W., \& 
Garmire, G. P. 2004, \apj, 603, 548

\bibitem[]{2040}Pierce-Price, D.,  Richer, J. S.,  Greaves, J. S.,  
Holland, W. S.,  Jenness, T., 
Lasenby, A. N. et al. 2000, ApJ, 545, 121

\bibitem[]{2044}
Reich, W. 2003, A\&A, 401, 1023

\bibitem[Rec]{rev04} Revnivtsev, M. G.,  Churazov, E. M.,  Sazonov, S. Yu.,  Sunyaev, R.A.,  
Lutovinov, A. A. et al. 2004, A\&A, 425, 49

\bibitem[]{2050}
Rodriguez-Fernandez, N.J., Martin-Pintado, J., Fuente, A., de Vicente, P., Wilson, T.L. \& 
H\"uttemeister 2001, A\&A, 365, 174

\bibitem[]{2054}
Rodriguez-Fernandez, N.J.,  Martin-Pintado, J., Fuente, A. \& Wilson, T.L. 2004, 
A\&A, 427,  217


\bibitem[]{2059} 
Roy, S. 2003, A\&A, 403, 917

\bibitem[]{2062} 
Rudolph, A. L.,  Fich, M.,  Bell, G. R.,  Norsen, T., Simpson, J., P., 
 Haas, M. R. \& Erickson, E. F  2006, ApJS, 162, 346	


\bibitem[]{2067}
Sakano, M., Warwick, R.S., Decourchelle, A. \& Predhel, P. 2003,
MNRAS, 340, 747

\bibitem[]{2071}
Serabyn, E. \& G\"usten, R. 1987, A\&A, 184, 133

\bibitem[]{2074} Sunyaev, R.A., Markevitchm M. \& Pavlinsky, M. 1993, ApJ, 
407, 606

\bibitem[]{2077} Tatischeff, V. 2003,
in Final stages of stellar evolution,
ed. C. Motch \& J.-M. Hameury, EAS Pub Ser, 79

\bibitem[Townsley \etal(2002b)]{tow02b} Townsley, L. K. \etal, 
2002b, 
  NIM-A, 486, 751

\bibitem[]{2085}
Tsuboi, M., Kobayashi, H., Ishiguro, M. \&  Murata, Y. 1991, PASJ, 43, L27
 
\bibitem[]{2088} Tsuboi, M., M., Ukita, N., \& Handa, T. 1997, ApJ, 481,
263


\bibitem[]{2092}
Valinia, A. \& Marshall, F.E. 2000, ApJ, 505, 134


\bibitem[]{2096}
van der Tak, F.F.S., Belloche, A., Schilke, P., G\"usten, R., Philipp, S., Comito, et al. 2006, 
A\&A, in press

\bibitem[]{2100}
Wang, Q.D., Dong, Hui \& Lang, C. 2006, ApJ (in press)

\bibitem[Wang \etal(2002)]{wgl02} Wang, Q. D., Gotthelf, E. V., \& 
   Lang, C. C. 2002, \nat, 415, 148


\bibitem[Weisskopf \etal(2002)]{wei02} Weisskopf, M. C., Brinkman, B., 
  Canizares, C., Garmire, G., Murray, S., \& van Speybroeck, L. P. 2002, 
  \pasp, 114, 1

\bibitem[]{2111}
Wilson, T.L. \& Rood, R. 1994,  ARAA 32, 19

\bibitem[Yusef-Zadeh(2003)]{yusef-zadeh03} Yusef-Zadeh, F., 2003, ApJ, 598, 325

\bibitem[]{2116}
Yusef-Zadeh, F. 1986, PhD thesis, Columbia University


\bibitem[]{2120}
Yusef-Zadeh, F., Hewitt, J. W., \& Cotton, W. 2004, ApJS, 155, 421

\bibitem[]{2123} 
Yusef-Zadeh, F., Law, C. \& Wardle 2002, ApJ, 568, L121

\bibitem[Yusef-Zadeh, Morris \& Chance(1984)]{yusef-zadeh84} Yusef-Zadeh, F., Morris, M, \& 
Chance, D., 1984, Nature, 310, 557

\bibitem[]{2129}
Yusef-Zadeh, F., Morris, M., Slee, O.B. \& Nelson, G.J. 1986, ApJ, 
310, 689

\bibitem[]{2133}
Yusef-Zadeh, F., Nord, M., Wardle, M., Law, C., Lang, C. \& Lazio, T.J.W. 2003, 
ApJ, 590, L103
 
\bibitem[]{2137}
Yusef-Zadeh, F.; Wardle, M.; Muno, M.; Law, C.; Pound, M. 2005, 
AdSR, 35, 1974





 

\end{thebibliography}
\end {document}